\documentclass[aps,prd,reprint,preprintnumbers,nofootinbib]{revtex4-1}
\pdfoutput=1

\usepackage[utf8]{inputenc}
\usepackage{amsmath}
\usepackage{amsfonts}
\usepackage{amssymb}
\usepackage{color}
\usepackage{graphicx}
\usepackage{bbold}
\usepackage{mathtools}

\usepackage{bbm}
\usepackage{hyperref}
\usepackage{bm}

\newcommand{\orcid}[1]{\href{https://orcid.org/#1}{#1}}

\DeclareMathOperator{\diag}{diag}
\DeclareMathOperator{\sign}{sign}

\begin{document}

\title{Parameter symmetries of neutrino oscillations in vacuum, matter,\\ and approximation schemes}


\author{Peter B.~Denton}
\thanks{pdenton@bnl.gov,  orcid \# \orcid{0000-0002-5209-872X}}
\affiliation{High Energy Theory Group, Physics Department, Brookhaven National Laboratory, Upton, NY 11973, USA}

\author{Stephen J.~Parke}
\thanks{parke@fnal.gov,  orcid \# \orcid{0000-0003-2028-6782}}
\affiliation{Theoretical Physics Dept., Fermi National Accelerator Laboratory, Batavia, IL 60510, USA}


\date{December 8, 2021}

\newcommand{\Dmsqee}{\Delta m^2_{ee}}
\newcommand{\eps}{\epsilon}
\newcommand{\bra}[1]{\langle#1|}
\newcommand{\ket}[1]{|#1\rangle}
\newcommand{\then}{\Rightarrow}
\newcommand{\td}[2]{\frac{d#1}{d#2}}

\newcommand{\cc}{c_{(\phi-\theta_{13})}}
\newcommand{\s}{s_{(\phi-\theta_{13})}}
\newcommand{\Dl}[2]{\Delta\lambda_{#1#2}}
\newcommand{\e}[1]{\times10^{#1}}
\newcommand{\wh}[1]{\widehat{#1}}
\newcommand{\wt}[1]{\widetilde{#1}}

\begin{abstract}
Expressions for neutrino oscillations contain a high degree of symmetry, but typical forms for the oscillation probabilities mask these symmetries  of the oscillation parameters. We elucidate the $2^7$
 parameter symmetries of the vacuum parameters and draw connections to the choice of definitions of the parameters as well as interesting degeneracies.
We also show that in the presence of matter an \emph{additional} set of $2^7$  parameter symmetries exist of the matter parameters.
Due to the complexity of the exact expressions for neutrino oscillations in matter,  numerous approximations have been developed; we show that under certain assumptions, approximate expressions have at most $2^6$  
additional  parameter symmetries of the matter parameters.
We also include one parameter symmetry related to the LMA-Dark degeneracy that holds under the assumption of CPT invariance; this adds one additional factor of two to all of the above cases.
Explicit, non-trivial examples are given of how physical observables in neutrino oscillations, such as the probabilities, CP violation, the position of the solar and atmospheric resonance, and the effective $\Delta m^2$'s for disappearance probabilities, are invariant under all of the above symmetries.
We investigate which of these parameter symmetries apply to numerous approximate expressions in the literature and show that a more careful consideration of symmetries improves the precision of approximations.
\end{abstract}


\preprint{FERMILAB-PUB-21-279-T, arXiv:2106.12436}

\maketitle


\section{Introduction}
The propagation of neutrinos is described by the eigenvalues and eigenvectors of the Hamiltonian.
The eigenvectors form up into a unitary matrix which, after rephasing of the charged leptons and the neutrinos\footnote{Neutrinos can be rephased if they have only Dirac mass terms or are in the ultra-relativisitic $p \gg m$ regime {; since we are focused on oscillations the latter condition always holds}.} results in four degrees of freedom.
There are a considerable number of options for parameterizing the matrix, see e.g.~\cite{Denton:2020igp} for a recent overview.
Even within one parameterization scheme, there may still be a large number of choices to be made, and that is the focus of this paper.

The lepton\footnote{Many of the points made here apply to the quark mixing matrix \cite{Cabibbo:1963yz,Kobayashi:1973fv} as well, however there is no connection  {to} the Majorana phases and the matter effect isn't relevant.} mixing matrix \cite{Pontecorvo:1957qd,Maki:1962mu} is usually described \cite{Zyla:2020zbs} as the product of three rotations: (23), (13), (12), with a complex phase associated with the (13) rotation resulting in four different physical degrees of freedom: $\theta_{23}$, $\theta_{13}$, $\theta_{12}$, and $\delta$.
 {Distilling the 18 degrees of freedom of a complex $3\times3$ matrix down to four parameters is due to the fact that the mixing matrix is unitary\footnote{ {We assume that any deviations from unitarity due to sterile neutrino hints or neutrino mass generation are negligible.}}, but any choice of parameterization as a sequence of rotations leaves a number of discrete symmetries of the parameters including the usual PDG choice. 
Once the order of rotations is chosen and the parametrization is made,} there are still many different configurations that result in the same physics.
The  {parameter} symmetries that connect these different configurations are directly related to the definition of the range of the parameters {, see e.g.~\cite{Gluza:2001de}}.
Typically the initial definition made is to restrict the ranges of the three mixing angles from $[0,2\pi)$ down to $[0,\pi/2)$ - a factor of four reduction for each angle.
That implies that each quadrant of each angle can be related to the other quadrants, implying $4^3=64$ discrete  {parameter} symmetries.
This means that for any given configuration of the mixing angles, there are 63 others that lead to exactly equivalent physics.
There is  {also} one discrete  {parameter} symmetry which again doubles the number of  {parameter} symmetries related to the choice of how one defines the mass eigenstates \cite{Denton:2020exu}. 
This leads to a total of $2^7=128$ discrete  {parameter} symmetries.
 {Ref.~\cite{Gluza:2001de} concentrated on the allowed range for each parameter, where in this paper we concentrate on the non-trivial parameter symmetries of the physical oscillation observables when written in terms of mixing angles and a phase.
These parameter symmetries restrict the allowed combinations of such parameters which we elucidate with important examples.}
 {In addition, ref.~\cite{Gonzalez-Garcia:2011vlg} assumed CP conservation and vanishing $\Delta m^2_{21}$ and examined a subset of the symmetries presented in this work to determine the allowed ranges of the parameters.}
In addition, if one generalizes the mixing matrix to include a phase on each rotation, there are two additional continuous degrees of freedom related to rephasing of the neutrino states or the Majorana phases.
In this paper we show exactly how all of the  {parameter} symmetries arise.

Going beyond the symmetries of the vacuum parameters which can all be addressed via a choice of definition of the mixing angles and mass eigenstates, these  {parameter} symmetries also naturally extend to the mixing angles and eigenvalues in matter as well, providing another $2^7=128$ discrete  {parameter} symmetries.
Since both of these sets of symmetries apply simultaneously in matter, there are in total $2^{14}=16,384$ discrete  {parameter} symmetries of the oscillation probabilities in matter  {for both neutrinos and anti-neutrinos}.

 {The above symmetries all leave the Hamiltonian unchanged up to rephasing.
There is an additional parameter symmetry which changes the Hamiltonian but leaves all physical observables unchanged known as the CPT parameter symmetry.
This is due to the fact that, under the assumption that CPT is a good symmetry, all of the oscillation physics is unchanged by the transformation $H\to-H^*$.
This symmetry, when combined with some of the others discussed above, is equivalent to the so-called LMA-Light -- LMA-Dark symmetry often discussed in the literature \cite{deGouvea:2000pqg,Miranda:2004nb,Bakhti:2014pva,Coloma:2016gei,Denton:2018xmq}.
This leads to an additional power of two for the total number of symmetries for each of the vacuum case, the matter case, and approximate case.}

It is well known that the exact solutions to neutrino oscillations in constant matter density are quite intractable.
To gain insights on the physics of neutrino oscillations in matter, a large number of approximate expressions have been developed.
We also examine these  {parameter} symmetries in the scenario of a perturbative Hamiltonian.
The symmetries of both the vacuum parameters and the new approximate parameters also apply subject to certain conditions, one of which restricts the number of  {parameter} symmetries of the perturbative parameters by two to bring the total number of  {parameter} symmetries for a perturbative scheme to $2^{13}=8,192$.
Finally, we examine which of these  {parameter} symmetries are satisfied by various approximate expressions in the literature.

\section{ {Parameter} symmetries}
\label{sec:symmetries}
Since the Hamiltonian in the flavor basis uniquely and exactly determines the neutrino oscillation probabilities, any  {parameter} symmetry of this Hamiltonian is necessarily a  {parameter} symmetry of the probabilities.
We first focus on the vacuum case for simplicity, and will later show that the presence of matter, handled either exactly or perturbatively, behaves in a similar way.
We define the Hamiltonian\footnote{We only focus on the flavor basis in this paper.} in the usual PDG fashion \cite{Zyla:2020zbs} except with a complex phase on each rotation,
\begin{equation}
H_{\rm flav} =\frac1{2E}U_{23}U_{13}U_{12}M^2U_{12}^\dagger U_{13}^\dagger U_{23}^\dagger\,,
\label{eq:Hvac}
\end{equation}
where the  {nontrivial part of the} $2\times2$ submatrix of the $3\times3$ complex rotation matrices are defined as\footnote{We use the usual $s_{ij}=\sin\theta_{ij}$, $c_{ij}=\cos\theta_{ij}$, $\Delta m^2_{ij}=m^2_i-m^2_j$ convention.}
\begin{equation}
U_{ij}(\theta_{ij},\delta_{ij})=
\begin{pmatrix}
c_{ij}&s_{ij}e^{i\delta_{ij}}\\
-s_{ij}e^{-i\delta_{ij}}&c_{ij}
\end{pmatrix}\,,
\end{equation}
and the mass-squared matrix is $M^2=\diag(m_1^2,m_2^2,m_3^2)$.
That is, we describe the mixing matrix with three rotation angles $\theta_{ij}$ and three associated complex phases $\delta_{ij}$.
To understand the  {parameter} symmetries it is useful to work with this slightly more general mixing matrix with three distinct complex phases, one for each rotation, however ultimately oscillation probabilities only depend on the sum of these three complex phases.

\subsection{Discrete  {parameter} symmetries}
We now list the various  {parameter} symmetries of the Hamiltonian.
First, there are 128 discrete  {parameter} symmetries of the vacuum Hamiltonian  {up to rephasing}.
To describe these we introduce the parameters $m_{ij}$ and $n_{ij}$ each of which are $\in\mathbb Z_2$ (that is, $\{0,1\}$).
A sign flip of a cosine term is given by the $m_{ij}$ parameters: $c_{ij}\to(-1)^{m_{ij}}c_{ij}$ and a sign flip of a sine term is given by the $n_{ij}$ parameters: $s_{ij}\to(-1)^{n_{ij}}s_{ij}$.
This allows us to write down ``angle reflection'' ($n_{ij}$) and ``angle shift and reflection'' ($m_{ij}$)  {parameter} symmetries as follows:
\begin{itemize}
\item $(i,j)=(2,3)$ or $(1,2)$:
\begin{align}
&c_{ij}\to(-1)^{m_{ij}}c_{ij}\,,\ s_{ij}\to(-1)^{n_{ij}}s_{ij} \notag \\[2mm]  &\text{ so long as }\delta_{ij}\to\delta_{ij}+(m_{ij}+n_{ij})\pi\,.
\end{align}
\item $(i,j)=(1,3)$:
\begin{align}
&c_{13}\to(-1)^{m_{13}}c_{13}\,,\ s_{13}\to(-1)^{n_{13}}s_{13}   \notag \\[2mm] & \text{ so long as }\delta_{13}\to\delta_{13}+n_{13}\pi\,.
\end{align}
\end{itemize}
In each case, if one of the angles is changed via the $m_{ij}$ or $n_{ij}$ parameters then the corresponding $\delta_{ij}$ must accumulate a factor of $\pi$ with the exception of changes to $c_{13}$.
That is, the six $m_{ij}$ and $n_{ij}$ are all independent and can each be either zero or one, leading to $2^6=64$ combinations.

The remaining discrete  {parameter} symmetry is the 1-2 interchange \cite{Denton:2016wmg} for which all of the following simultaneously happen,
\begin{itemize}
\item $(i,j)=(1,2)$:
\begin{align}
& c_{12}\to s_{12}\text{ and }s_{12}\to c_{12}  \notag \\[2mm] & \text{ so long as }\delta_{12}\to\delta_{12}+\pi\text{ and }m_1^2\leftrightarrow m_2^2\,.
\label{eq:12interchange}
\end{align}
\end{itemize}
The final part of the interchange is equivalent to $\Delta m^2_{21}\to\Delta m^2_{12}=-\Delta m^2_{21}$ and $\Delta m^2_{31}\leftrightarrow\Delta m^2_{32}$.
Note that the 1-2 interchange can be done simultaneously with the shift and reflection described above\footnote{While the $m_{ij}$ and $n_{ij}$ notation makes it appear as though the 1-2 interchange does not commute with a  {parameter} symmetry related to either $m_{12}$ or $n_{12}$, note that $m_{12}$ and $n_{12}$ always appear as a sum, so swapping one for the other leads to no change.}.
The  {parameter} symmetry related to this interchange provides one more factor of two leading to 128 total  {parameter} symmetries.

After performing any combination of these  {parameter} symmetries including the 1-2 interchange, the Hamiltonian is exactly the same up to a possible overall phase matrix,
\begin{align}
& H_{\rm flav}\to
\label{eq:Htransformation}
  \\ & \diag((-1)^{m_{13}+m_{23}},1,1)H_{\rm flav}\diag((-1)^{m_{13}+m_{23}},1,1)\,, \notag
\end{align}
see appendix \ref{sec:transformation} for an explicit derivation of this.
This $(-1)$ phase coming from $m_{13}$ and/or $m_{23}$ terms can then be absorbed into the definition of $\ket{\nu_e}$ and nothing has changed.
The presence of $m_{13}$ and $m_{23}$ and not $m_{12}$ is due to the fact that the sign from sending $c_{12}\to-c_{12}$ can go through the middle of the Hamiltonian and commutes with the $M^2$ matrix, while this isn't true for the $m_{13}$ or $m_{23}$  {parameter} symmetries.
The $n_{ij}$  {parameter} symmetries operate on the rotation matrix level and thus do not affect the total Hamiltonian.

The various interchanges can be thought of equally in terms of sines and cosines (e.g.~$s\to-s$) or in terms of angles (e.g.~$\theta\to-\theta$).
For convenience we include the relationship between these in a table:
\begin{center}
\begin{tabular}{l|l}
sines/cosines&angles\\\hline
$s\to-s$, $c\to c$&$\theta\to-\theta$\\
$s\to s$, $c\to-c$&$\theta\to\pi-\theta$\\
$s\to-s$, $c\to-c$&$\theta\to\pi+\theta$\\\hline
$c\to s$, $s\to c$&$\theta\to\pi/2-\theta$\\
$c\to-s$, $s\to c$&$\theta\to\pi/2+\theta$\\
$c\to s$, $s\to-c$&$\theta\to-\pi/2+\theta$\\
$c\to-s$, $s\to-c$&$\theta\to-\pi/2-\theta$\\
\end{tabular}
\end{center}

\subsection{Continuous  {parameter} symmetries}
In addition to the above discrete  {parameter} symmetries, there are two continuous degrees of freedom as well dubbed the ``delta shuffle''\footnote{These were pointed out in \cite{Rodejohann:2011vc}.}.
We define the three rotations and complex phases by\footnote{The choice of a minus sign on $\delta_{13}$ is arbitrary and is chosen for consistency with the PDG convention.}
\begin{equation}
U_{23}(\theta_{23},\delta_{23})U_{13}(\theta_{13},-\delta_{13})U_{12}(\theta_{12}, \delta_{12})\,.
\label{eq:Udeltas}
\end{equation}
In fact, in the context of neutrino oscillations, these three phases are related by
\begin{equation}
\delta_{23}+\delta_{13}+\delta_{12}=\delta\mod\ 2\pi\,,
\end{equation}
where $\delta$ is a new fixed parameter and is equal to the usual single complex phase.
That is, there are two additional degrees of freedom which, it turns out, are equivalent to the Majorana phases, see section \ref{sec:discussion}.
For example, the three following scenarios, each with a single complex phase, are all equivalent,
\begin{gather}
U_{23}(\theta_{23},\delta)U_{13}(\theta_{13}, 0)U_{12}(\theta_{12}, 0)\,,\label{eq:Udeltas23}\\
U_{23}(\theta_{23},0)U_{13}(\theta_{13}, -\delta)U_{12}(\theta_{12}, 0)\,,\label{eq:Udeltas13}\\
U_{23}(\theta_{23},0 )U_{13}(\theta_{13}, 0)U_{12}(\theta_{12},\delta)\,.\label{eq:Udeltas12}
\end{gather}
See appendix \ref{sec:shuffle explicit} for the explicit rephasing expression relating eq.~\ref{eq:Udeltas} to eqs.~\ref{eq:Udeltas23}-\ref{eq:Udeltas12}.
This can also be expressed from the point of view of rephasing the flavor states as well as the mass eigenstates,
\begin{equation}
D_fU_{23} U_{13} U_{12} D_m M^2 D_m^\dagger U_{12}^\dagger U_{13}^\dagger U_{23}^\dagger D_f^\dagger\,,
\end{equation}
where $D_f$ and $D_m$ are diagonal rephasing matrices, $\diag(e^{i\alpha},e^{i\beta},e^{i\gamma})$.
Note that since $D_m$ commutes with $M^2$, it trivially cancels while the $D_f$ rephasing can be absorbed into the definitions of the neutrino  {flavor} states.
The $D_f$ rephasing matrix requires some care if the Hamiltonian is split into two parts for a perturbative description; see section \ref{sec:perturbative} below.

\subsection{ {CPT parameter symmetry}}
\label{sec:cpt}
 {All of the above parameter symmetries are symmetries of the vacuum Hamiltonian up to rephasing, $H\to D_fHD_f^\dagger$ where $D_f$ is an arbitrary rephasing matrix.
There is an additional parameter symmetry of all oscillation observables (e.g.~the probabilities) that is not of the above form that depends on the assumption of CPT invariance \cite{deGouvea:2000pqg,Minakata:2002qe}.
This parameter symmetry can be written as,
\begin{equation}
m_i^2\to-m_i^2\ \text{and}\ \delta\to-\delta\,.
\end{equation}
Note that this requires the sum of the three phases, $\sum\delta_{ij}$, to change signs.
This is a parameter symmetry of the vacuum Hamiltonian because it is equivalent to sending $H\to-H^*$.
The minus sign applies a time reversal to the vacuum Hamiltonian and the complex conjugate applies a charge-parity reversal to the vacuum Hamiltonian; under the reasonable assumption that CPT is a good symmetry at scales relevant for neutrino oscillations, all physical observables remain the same.}

\subsection{Summary of parameter symmetries}
\label{sec:summary of symmetries}
These  {parameter} symmetries can be rewritten relatively compactly.
For $m_{ij},n_{ij}\in\{0,1\}$, the following are exact  {parameter} symmetries of the Hamiltonian:
\begin{itemize}
\item $(13)$: $c_{13} \rightarrow (-1)^{m_{13}} c_{13}$, $s_{13} \rightarrow (-1)^{n_{13}} s_{13}$, \\[2mm] and $\delta_{13} \rightarrow \delta_{13} \pm n_{13}\pi $,
\item $(23)$: $c_{23} \rightarrow (-1)^{m_{23}} c_{23}$, $s_{23} \rightarrow (-1)^{n_{23}} s_{23}$, \\[2mm] and $\delta_{23} \rightarrow \delta_{23} \pm (m_{23}+n_{23}) \pi $,
\item $(12)$: $c_{12} \rightarrow (-1)^{m_{12}} c_{12}$, $s_{12} \rightarrow (-1)^{n_{12}} s_{12}$, \\[2mm] and $\delta_{12} \rightarrow \delta_{12} \pm (m_{12}+n_{12}) \pi$,
\item $(12)$: $c_{12} \rightarrow (-1)^{m_{12}} s_{12}$, $s_{12} \rightarrow (-1)^{n_{12}} c_{12}$, \\[2mm] and $\delta_{12} \rightarrow \delta_{12} \pm (m_{12}+n_{12}+1) \pi $ plus $m_1^2\leftrightarrow m_2^2$,
\item For three phases defined:\\[2mm]  $U_{23} (\theta_{23},\delta_{23}) U_{13} (\theta_{13},-\delta_{13}) U_{12}(\theta_{12},\delta_{12})$,\\[2mm] there are two free degrees of freedom after applying the constraint: $\delta_{23}+\delta_{13}+\delta_{12}=\delta$.
\item  {CPT: $m_i^2\to-m_i^2$ and $\delta\to-\delta$,}
\end{itemize}
See, for example, \cite{Parke:2018shx} for neutrino oscillation amplitudes in vacuum which explicitly satisfy all of these discrete  {parameter} symmetries without additional manipulation, up to an overall unphysical phase.

\section{ {Parameter} symmetries in different Hamiltonians}
While the previous discussion was focused on the vacuum Hamiltonian, we now show that it extends to the matter Hamiltonian for constant (e.g.~long-baseline accelerator), smoothly varying (day-time solar), or sharply varying (atmospheric and night-time solar) density profiles.
In addition, it also applies to both exact and perturbative scenarios, provided that the perturbative approach satisfies certain properties discussed in subsection \ref{sec:perturbative} below.

The three forms of the Hamiltonian considered in this paper are,
\begin{align}
H_{\rm flav} ={}&\frac1{2E}\left[U_{23} U_{13} U_{12}M^2U_{12}^\dagger U_{13}^\dagger U_{23}^\dagger +A\right]\,, \label{eq:Hmat}\\
={}&\frac1{2E}W_{23} W_{13} W_{12}\Omega W_{12}^\dagger W_{13}^\dagger W_{23}^\dagger\,, \label{eq:Hmatdiag}\\
={}&\frac1{2E}\left[V_{23} V_{13} V_{12}\Lambda V_{12}^\dagger V_{13}^\dagger V_{23}^\dagger\right.  \label{eq:Hmatpert}\\
 & \hspace*{-1.5cm} \left.+ (U_{23} U_{13} U_{12}M^2U_{12}^\dagger U_{13}^\dagger U_{23}^\dagger + A -V_{23} V_{13} V_{12} \Lambda V_{12}^\dagger V_{13}^\dagger V_{23}^\dagger )\right]\,.   \notag 
\end{align}
each of which are exactly equivalent  {and all terms have identical mathematical structure of the form $U M^2 U^\dagger$ apart from matter term.}
In general the rotation matrices have complex phases associated with them as in the vacuum case, see eq.~\ref{eq:Udeltas}.
The rotation and diagonal matrices are as follows: ($U,M^2$) in eq.~\ref{eq:Hmat} are the vacuum parameters and ($W,\Omega$) in eq.~\ref{eq:Hmatdiag} are the diagonalized exact eigenvectors and eigenvalues in matter \cite{Zaglauer:1988gz}.
For eq.~\ref{eq:Hmatpert} ($V,\Lambda$) are any approximation\footnote{These approximations need not be good approximations for these  {parameter} symmetries to hold.} for the Hamiltonian in matter that is diagonalized by rotations of the order: (23), (13), (12) of any angles and phases, e.g.~DMP \cite{Denton:2016wmg}.
That is, some of the $V$ could be vacuum rotations ($U$), exact rotations ($V$), or anything else.
 {We use hats to denote the parameters that exactly diagonalize the Hamiltonian as shown in eq.~\ref{eq:Hmatdiag}, e.g.~$\wh\theta_{ij}$ and $\wh\delta_{ij}$.}
We use tildes to denote the parameters of the first line of the perturbative Hamiltonian in eq.~\ref{eq:Hmatpert}, e.g.~$\wt\theta_{ij}$ and $\wt\delta_{ij}$ and the indices for the  {parameter} symmetries are similarly expressed as $\wt m_{ij}$ and $\wt n_{ij}$.
Typically the first line of eq.~\ref{eq:Hmatpert} is considered $H_0$ and the second line $H_1$.

 {Since we are working at the Hamiltonian level, any symmetry of the Hamiltonian is necessarily a symmetry of the oscillation probability regardless of whether the matter density profile is constant or not.
For example, if the density profile is a complicated function such as for atmospheric or solar neutrinos, it may be difficult to solve the Schr\"odinger equation (analytically or numerically), but the symmetries are still valid since all of the information required for propagation is in the Hamiltonian.
In this case the effective oscillation parameters in matter, $\wh\theta_{ij}$, $\wh\delta_{ij}$, and $\Delta\wh{m^2}_{ij}$, all evolve as well, but since the symmetries discussed below do not depend on the value of the matter potential, $a$, they all apply for the entire probability for any matter density profile\footnote{ {Note that these symmetries for supernova neutrinos, where neutrino-neutrino interactions are relevant, require additional care.}}.}

\subsection{ {Parameter} symmetries in matter}
Since $A\equiv\diag(a,0,0)$ the matter effect matrix\footnote{The matter effect is given by $a=2\sqrt2G_FEN_e\rho$ \cite{Wolfenstein:1977ue}.} respects all of the  {parameter} symmetries in question  {apart from the CPT parameter symmetry}, adding $A$ to the vacuum Hamiltonian in eq.~\ref{eq:Hmat} results in a matrix that also respects the  {parameter} symmetries when described in terms of the vacuum parameters.
Therefore the diagonalization of the Hamiltonian in matter also makes no difference since the Hamiltonians in eqs.~\ref{eq:Hmat} and \ref{eq:Hmatdiag} are equal.
This means that eq.~\ref{eq:Hmatdiag} also respects the  {parameter} symmetries in terms of the vacuum parameters.

Moreover, since it is now written in the same form as that of the vacuum Hamiltonian, the new diagonalized parameters (the eigenvalues, the mixing angles and complex phase of the eigenvectors in matter)  {also obey an additional set of parameter} symmetries  {summarized here.
For $\wh{m}_{ij},\wh{n}_{ij}\in\{0,1\}$:
\begin{itemize}
\item $(13)$: $\wh{c}_{13} \rightarrow (-1)^{\wh{m}_{13}} ~ \wh{c}_{13}$, $\wh{s}_{13} \rightarrow (-1)^{\wh{n}_{13}} ~ \wh{s}_{13}$, \\[2mm] and $\wh{\delta}_{13} \rightarrow \wh{\delta}_{13} \pm \wh{n}_{13}\pi $,
\item $(23)$: $\wh{c}_{23} \rightarrow (-1)^{\wh{m}_{23}} ~\wh{c}_{23}$, $\wh{s}_{23} \rightarrow (-1)^{\wh{n}_{23}} ~\wh{s}_{23}$, \\[2mm] and $\wh{\delta}_{23} \rightarrow \wh{\delta}_{23} \pm (\wh{m}_{23}+\wh{n}_{23}) \pi $,
\item $(12)$: $\wh{c}_{12} \rightarrow (-1)^{\wh{m}_{12}} ~\wh{c}_{12}$, $\wh{s}_{12} \rightarrow (-1)^{\wh{n}_{12}} ~\wh{ s}_{12}$, \\[2mm] and $\wh{\delta}_{12} \rightarrow \wh{\delta}_{12} \pm (\wh{m}_{12}+\wh{n}_{12}) \pi$,
\item $(12)$: $\wh{c}_{12} \rightarrow (-1)^{\wh{m}_{12}} ~\wh{s}_{12}$, $\wh{s}_{12} \rightarrow (-1)^{\wh{n}_{12}} ~\wh{c}_{12}$, \\[2mm] and $\wh{\delta}_{12} \rightarrow \wh{\delta}_{12} \pm (\wh{m}_{12}+\wh{n}_{12}+1) \pi $ \\[1.5mm] 
plus $\wh{{m^2}_1}\leftrightarrow \wh{{m^2}_2}$,
\item For three phases defined: \\[2mm] $W_{23} (\wh{\theta}_{23},\wh{\delta}_{23}) W_{13} (\wh{\theta}_{13},-\wh{\delta}_{13}) W_{12}(\wh{\theta}_{12},\wh{\delta}_{12})$, \\[2mm] there are two free degrees of freedom after applying the constraint: $\wh{\delta}_{23}+\wh{\delta}_{13}+\wh{\delta}_{12}=\wh{\delta}$.
\end{itemize}
}

 {Therefore,} the exact expressions for neutrino oscillations in matter \cite{Zaglauer:1988gz,Kimura:2002wd,Denton:2019ovn} respect the 128 symmetries of the vacuum parameters as described  {in the previous section except for the CPT parameter symmetry}.
They also respect 128 symmetries in terms of the matter parameters described above, for a total of $2^{14}=16,384$ possible  {vacuum plus matter parameter} symmetries.  {This includes} flipping $\ket{\nu_1}\leftrightarrow\ket{\nu_2}$ and/or $\ket{\wh\nu_1}\leftrightarrow\ket{\wh\nu_2}$ where the $\ket{\wh\nu_i}$ are the exact eigenstates of the Hamiltonian in matter.  {Adding the CPT parameter symmetry, given by\footnote{ {The CPT parameter symmetry of the matter parameters can be achieved by  {the simultaneous transformations} $m_i^2\to-m_i^2$, $\delta\to-\delta$, and $a\to-a$.}}
\begin{itemize}
\item CPT: $\wh{m_i^2}\to-\wh{m_i^2}$ and $\wh{\delta}\to-\wh{\delta}$\,,
\end{itemize}
doubles the total number of symmetries. All of these symmetries impose constraints on the analytic form of the oscillation probabilities matter as will be discussed in section \ref{sec:discussion}.}

Each of these parameter symmetries can be applied at each point in propagation in either a constant or varying matter potential and the physics at the step remains invariant. In principle, one could choose a different parameter symmetry at each step, however, for continuity of the mixing angles and CP phase, applying the same parameter symmetries along the entire route is certainly the simplest option and the physics cannot depend on the choice made.

\subsection{ {Parameter} symmetries of a perturbative Hamiltonian}
\label{sec:perturbative}
Since the exact expressions for neutrino oscillations in matter for constant or sharply varying density profiles tend to be fairly opaque, numerous approximation schemes have been considered in the literature, for an overview see \cite{Parke:2019vbs}.
One technique is that of splitting the Hamiltonian into a large part and a small part: $H=H_0+H_1$.
$H_0$ is  {then} the diagonal part of $H$ after successive diagonalizing with two component rotations until the off-diagonal elements are sufficiently small.
Then, $H_1$ is just the remaining off-diagonal part, see \cite{Agarwalla:2013tza,Minakata:2015gra,Denton:2016wmg,Denton:2018fex,Denton:2019qzn}.
While different  {perturbative schemes} have different benefits\footnote{If in the approximation scheme you are considering, $V_{12}= \mathbb I$, such as \cite{Minakata:2015gra}, then a $\ket{\wh{\nu_1}}\leftrightarrow\ket{\wh{\nu_3}}$ interchange symmetry is possible which can be made exact by appropriate choices for $\wh{\theta}$ and $\wh{\delta}$ in $V_{13}$, similar to what was performed for the $\ket{\wh{\nu_1}}\leftrightarrow\ket{\wh{\nu_2}}$ interchange symmetry in $V_{12}$.
Using the methods of this paper one can work out all the parameter symmetries for such approximation schemes, see appendix \ref{sec:MP}.}, we focus on that described in \cite{Denton:2016wmg} by Denton, Minakata, and Parke (DMP)  {as a concrete example}.

First, we note that the above  {parameter} symmetries still apply to the vacuum parameters, as they must, since the addition of the matter potential matrix is invariant under the  {parameter} symmetries.
Second, we find that there are four key conditions for a perturbative Hamiltonian to satisfy in order for the new approximate eigenvalues and approximate angles (denoted with tildes) to be independent under the  {parameter} symmetries.
\begin{enumerate}
\item The order of rotations must match that of the vacuum rotations.
\item The approximate eigenvalues must respect all the vacuum  {parameter} symmetries.
\item The phases of the new rotations must match the corresponding vacuum ones mod $\pi$,
\begin{equation}
\wt\delta_{ij}=\delta_{ij}\mod\ \pi\,.
\end{equation}
\item Certain vacuum and perturbative  {parameter} symmetries must match:
\begin{equation}
m_{13}+m_{23}=\wt m_{13}+\wt m_{23}\mod\ 2\,.
\label{eq:m13m23pert}
\end{equation}
\end{enumerate}
We now explain in more detail exactly what these conditions are.

First, it is necessary to follow the same sequence of rotations as in vacuum.
That is, for the PDG parameterization of the lepton mixing matrix \cite{Zyla:2020zbs}, the zeorth order part of the Hamiltonian must be diagonalized by a (23) rotation, then a (13) rotation, followed by a (12) rotation.
If the lepton mixing matrix is parameterized in a different way \cite{Denton:2020igp} then a different sequence of rotations would need to be used, which may or may not be advantageous depending on the exact region of interest.
This condition is satisfied in DMP \cite{Denton:2016wmg} (as well as in \cite{Minakata:2015gra}) where it was a noted convenient benefit that happened by chance due to focusing on the large off-diagonal elements and removing the level crossings.
Note that since the sequence of rotations used in \cite{Agarwalla:2013tza} is a different order, the approximate matter parameters used there do not satisfy these  {parameter} symmetries.

Secondly, in general, the approximate eigenvalues (these make up the $\Lambda$ matrix) need to respect the symmetries of the vacuum parameters in order for the  {parameter} symmetries to be satisfied by expressions derived from such an approximation scheme.
While the full Hamiltonian in eq.~\ref{eq:Hmatpert} does respect these  {parameter} symmetries, if the split between $H_0$ and $H_1$ is not done in a way that respects these  {parameter} symmetries, then expressions derived from only part of the Hamiltonian will not respect these vacuum  {parameter} symmetries.
We explicitly show in appendix \ref{sec:zs} that each of the eigenvalues in the exact solution respects the vacuum  {parameter} symmetries as they must, and in appendix \ref{sec:dmp} that each of the approximate eigenvalues in the DMP scheme do respect the vacuum  {parameter} symmetries.

Third, assuming the vacuum mixing matrix is parameterized with three phases as in eq.~\ref{eq:Udeltas}, the phases in the diagonalizing matrices $V_{ij}$ must be given by $\wt\delta_{ij}=\delta_{ij}\mod\ \pi$.
This condition guarantees that no net phase appears between the exact and approximate expressions.
Therefore there are no  {parameter} symmetry degrees of freedom related to the delta shuffle for the perturbative complex phases, $\wt\delta_{ij}$.
Note that in DMP this was satisfied as the vacuum matrix was parameterized with the phase on the (23) rotation as in eq.~\ref{eq:Udeltas23}, the (23) diagonalization matrix was the same as the vacuum parameters,
\begin{equation}
\sin2\wt\theta_{23}e^{i\wt\delta_{23}}=\sin2\theta_{23}e^{i\delta_{23}}\,.
\label{eq:23d def}
\end{equation}
This form of the definition is motivated as the imaginary part of eq.~\ref{eq:23d def} is known to be exactly satisfied for the exact versions of the matter variables \cite{Toshev:1991ku}\footnote{ {The Toshev identity, $\sin2\wt\theta_{23}\sin\wt\delta=\sin2\theta_{23}\sin\delta$ \cite{Toshev:1991ku}, gains a sign under certain changes in the definitions; this is consistent with the rest of the results of this paper since the Toshev identity is not a physically measurable quantity like $\Delta P_{\rm CP}$, given in eq.~\ref{eq:PCPV}.
With the correct signs, the Toshev identity reads $(-1)^{\wt m_{12}+\wt n_{12}+\wt n_{13}}\sin2\wt\theta_{23}\sin\wt\delta=(-1)^{m_{12}+n_{12}+n_{13}}\sin2\theta_{23}\sin\delta$ without $1 \leftrightarrow 2$ interchanges. With such interchanges in matter/vacuum an additional factor of (-1) is needed on matter/vacuum side of this generalized Toshev identity.}}.

Fourth, and most interestingly, certain combination of  {parameter} symmetries of the vacuum and perturbative parameters are not  {parameter} symmetries of the Hamiltonian.
Eq.~\ref{eq:Htransformation} shows how the Hamiltonian accumulates an overall phase if $m_{13}$ or $m_{23}$  {parameter} symmetries are applied.
As this phase can be absorbed into the definition of $\ket{\nu_e}$ it is not a problem, but if different phases appear on $H_0$ and $H_1$ then the phase cannot be simply absorbed into the definition of $\ket{\nu_e}$ anymore.
Since the impact of such a phase on $H_0$ comes only from the symmetries of the approximate parameters denoted $\wt m_{13}$ and $\wt m_{23}$ and the impact of such a phase on $H_1$ comes from not only $m_{13}$ and $m_{23}$ but also $\wt m_{13}$ and $\wt m_{23}$ as can be seen from the second line of eq.~\ref{eq:Hmatpert}, the only way to ensure that these phases can be absorbed into $\ket{\nu_e}$ is if the impact of both are the same.
This only happens when the conditions of eq.~\ref{eq:m13m23pert} are satisfied.
This extra condition implies that the number of symmetries in the perturbative parameters is a factor of 2 lower; one can think of it as once choices are made about $m_{23}$, $m_{13}$, and $\wt m_{23}$, then $\wt m_{13}$ is fixed.
Therefore there are an additional 128/2  {parameter} symmetries for the perturbative parameters for a total of $2^{13}=8,192$  {parameter} symmetries for a perturbative Hamiltonian that meets the above requirements.

\section{Discussion}
\label{sec:discussion}
\subsection{Definition of the parameters}
\label{sec:definition}
The $2^6=64$ discrete  {parameter} symmetries of the vacuum parameters not including the 1-2 interchange are exactly equivalent to the fact that one can define the range of each of the three mixing angles to be in only one quadrant.
That is, one could choose that each of the mixing angles $\theta_{ij}$ exists in $[\eta_{ij}\pi/2,(\eta_{ij}+1)\pi/2)$ where the $\eta_{ij}\in \mathbb Z$ need not be the same for each angle\footnote{In fact this generalizes somewhat.
The angles can be defined to be within any region given by $\theta\in\bigcup_i[x_i,y_i)$ for any $x_i$ and $y_i$ possibly different for each angle such that $\{|\cos\theta|\}$ are all unique as are $\{|\sin\theta|\}$ across the allowed range of $\theta$.
For example, one could define $\theta_{12}$ to be in the range $(-0.5\pi,-0.2\pi]\cup[0,0.2\pi)$.}.
For convenience, the standard convention is that $\eta_{ij}=0$ for each of the mixing angles.
This is equivalent to requiring that $s_{ij}\ge0$ and $c_{ij}\ge0$.

The remaining  {parameter} symmetry for the vacuum parameters is the 1-2 interchange symmetry.
Given the above range of the mixing angles and complete freedom in identifying the mass states, the 1-2 interchange symmetry exists.
As with the mixing angles, this implies that one should make a definition restricting this; there are two typical approaches to proceed (for a comprehensive examination of how one labels the mass states see e.g.~\cite{Denton:2020exu}).
First, one could fix the order of the mass states such that the first mass state is smaller than the second.
Second, one could fix $\theta_{12}$ to be contained within one octant, typically $\theta_{12}\in[0,\pi/4)$.
Each of these are equivalent as described by the interchange symmetry (up to a factor of $\pi$ on $\delta_{12}$).
Thus the fact that we can define $\Delta m^2_{21}>0$ \emph{or} $\theta_{12}\in[0,\pi/4)$ is exactly due to the 1-2 interchange symmetry.

We also note that for the same reason that there is a  {parameter} symmetry of mass states $\ket{\nu_1}$ and $\ket{\nu_2}$ related to $\theta_{12}$, there is also a different discrete  {parameter} symmetry related to flavor states $\ket{\nu_\mu}$ and $\ket{\nu_\tau}$ and $\theta_{23}$ for the same reason, also subject to appropriate modifications if the lepton mixing matrix is parameterized in a different way.
This  {parameter} symmetry is not a parameter symmetry like the others since we can differentiate between $\ket{\nu_\mu}$ and $\ket{\nu_\tau}$ by measuring the properties of their associated charged leptons.

\subsection{ {LMA-Dark degeneracy}}
 {
The CPT parameter symmetry discussed in section \ref{sec:cpt} has appeared in the literature typically accompanied with the 1-2 interchange parameter symmetry stated in eq.~\ref{eq:12interchange} above, and is often written as
\begin{equation}
c_{12}\leftrightarrow s_{12}\,,\ \Delta m^2_{31}\leftrightarrow-\Delta m^2_{32}\,,\ \text{and}\ \delta\to\pi-\delta\,.
\end{equation}
This form, or variations thereof, are often referred to as the LMA-Light -- LMA-Dark degeneracy or the Generalized Mass Ordering Degeneracy, see e.g.~\cite{Gonzalez-Garcia:2011vlg,Bakhti:2014pva,Coloma:2016gei,Denton:2018xmq}.
This particular combination of parameter symmetries is of interest due to interesting phenomenological implications, in particular when the matter effect is included.
In the presence of matter the  {vacuum} CPT parameter symmetry  {changes the Hamiltonian as follows;} 
\begin{equation}
H_{\rm vac}+A\to-H_{\rm vac}^*+A\,,
\label{eq:lmad}
\end{equation}
where $A\equiv\diag(a,0,0)$ and $a$ is the matter potential which is unchanged by this parameter symmetry.
This is the LMA-Light to LMA-Dark interchange.
The fact that this is the only parameter symmetry of the vacuum Hamiltonian but not of the matter Hamiltonian is exactly why measuring the matter effect (as has already been done by combining solar data \cite{SNO:2002tuh} with KamLAND data \cite{KamLAND:2013rgu}) is a necessary condition for measuring both mass orderings.

In the presence of new physics, it is possible that the matter effect could take the opposite sign of the expectation in the SM  {(i.e.~$A \rightarrow -A$)} where the new physics is described in the neutrino non-standard interactions (NSI) framework \cite{Wolfenstein:1977ue,Farzan:2017xzy,Dev:2019anc}.
This also means that, in the presence of NSIs, it is not possible to determine the atmospheric mass ordering since then even the matter Hamiltonian is invariant under $\Delta m^2_{31}\leftrightarrow-\Delta m^2_{32}$, although the details of the NSI model may allow one to break this degeneracy in most cases \cite{Gonzalez-Garcia:2011vlg,Coloma:2016gei,Denton:2018xmq}.

This parameter symmetry adds a factor of two to the number of symmetries in vacuum.
In matter it also adds a factor of two because, while the probabilities are no longer invariant under this parameter symmetry of the vacuum parameters as shown in eq.~\ref{eq:lmad}, they are invariant under the same symmetry but for the equivalent parameters in matter,
\begin{equation}
\wh{m_i^2}\to-\wh{m_i^2}\ \text{and}\ \wh\delta\to-\wh\delta\,.
\end{equation}
Similarly for any approximate diagonalization scheme, so long as all three eigenvalues change sign along with the complex phase, the CPT parameter symmetry holds.}

\subsection{Some technical details}
\label{sec:techdetails}
In this section we aim to understand these  {parameter} symmetries conceptually.
We begin with the delta shuffle since it is distinct from the others.
The main notable difference from the other  {parameter} symmetries is that this represents a continuous  {parameter} symmetry.
That is, it represents two additional continuous parameters in the matrix.
These two extra parameters are physical if neutrinos have a Majorana mass term \cite{Schechter:1980gk,Rodejohann:2011vc}.
If one writes the mixing matrix as a product of the usual PDG \cite{Zyla:2020zbs} form and the Majorana phase matrix $P=\diag(e^{-i\alpha},e^{-i\beta},1)$ as $U_{\rm PDG}P$, then our parameterization in eq.~\ref{eq:Udeltas} with $\delta_{ij}$ on each rotation is equivalent to $P^\dagger U_{\rm PDG}P$ which is equal to $U_{\rm PDG}P$ after rephasing the charged leptons.
The Majorana phases $\alpha$ and $\beta$ are related to the phases in our notation by,
\begin{align}
\delta&=\delta_{23}+\delta_{13}+\delta_{12}\,,\quad \quad
\alpha=\delta_{12}+\delta_{23}\,,\quad \quad
\beta=\delta_{23}\,.
\end{align}
See appendix \ref{sec:shuffle explicit} for more on rephasing.

Note that while the discrete  {parameter} symmetries are related to the definition of the ranges of the parameters (see section \ref{sec:definition}), they are still  {parameter} symmetries of the probabilities.
This means that every probability expression must respect these  {parameter} symmetries and they can be used as a valuable and quick cross check for any expression.

If the Hamiltonian is written as a series of three rotations in a different order, see \cite{Denton:2020igp}, the same  {parameter} symmetries apply, but care is required with regards to the inner vs.~outer rotations for the delta shuffle and the angle shift ($m_{ij}$).
It is the sign change on cosine in the middle rotation that does not result in a factor of $\pi$ on the associated complex phase and the 1-2 interchange applies only to the third rotation.
 {For example, if the lepton matrix is parameterized by a sequence of rotations in the order (23), (12), (13), then the 1-2 interchange would become a 1-3 interchange involving $\theta_{13}$ instead of $\theta_{12}$ and $m_1^2\leftrightarrow m_3^2$ instead of $m_1^2\leftrightarrow m_2^2$.
It would then be $\theta_{12}$ for which a $c_{12}\to-c_{12}$ change would not include a change to $\delta$, and it would be $m_{12}+m_{23}$ that would need to be equivalent between the vacuum and perturbative parameters.}

While the  {parameter} symmetries derived in the Hamiltonian framework presented in section \ref{sec:symmetries} are all exact, some care is required.
The Hamiltonian framework benefits from a high level of generality: any  {parameter} symmetries of the Hamiltonian are automatically  {parameter} symmetries of the physical observables - the probabilities.
It is possible, however, to artificially break the  {parameter} symmetries of the Hamiltonian while the  {parameter} symmetries of the probabilities persist.
For example, it is known that the probabilities are invariant under the addition of any matrix proportional to the identity matrix as this is equivalent to adding an overall phase which is not detectable in oscillations.
But if one adds a matrix that doesn't respect the  {parameter} symmetries, then the resultant Hamiltonian will also no longer respect some of the  {parameter} symmetries, but the probabilities still will of course.
For example, if one writes the Hamiltonian with $M^2=\diag(m_1^2,m_2^2,m_3^2)$ then all the  {parameter} symmetries are respected in the Hamiltonian.
If, however, one subtracts $m_1^2\mathbb I$ so that
\begin{equation}
M^2\to\diag(0,\Delta m^2_{21},\Delta m^2_{31})\,,
\label{eq:Msq0}
\end{equation}
the 1-2 interchange symmetry is not respected since $m_1^2\mathbb I$ is not invariant under the 1-2 interchange symmetry.
One can, however, subtract a term that is known to be invariant under the  {parameter} symmetries, see the previous paragraph, and the  {parameter} symmetries will still remain valid.
For example, one could subtract $(c_{12}^2m_1^2+s_{12}^2m_2^2)\mathbb I$ which sends
\begin{align}
M^2 &=\diag(m_1^2,m_2^2,m_3^2)  \notag  \\  &\to \diag(-s_{12}^2\Delta m^2_{21},c_{12}^2\Delta m^2_{21},\Dmsqee)\,,
\label{eq:Msqs12sq}
\end{align}
which still respects all the  {parameter} symmetries since the initial Hamiltonian does as does the subtracted identity matrix.
It is interesting to note then that these  {parameter} symmetries (particularly the 1-2 interchange symmetry) applies to the physical observables so long as there is at least one Hamiltonian which can be shifted to by a multiple of the identity matrix such that that  {parameter} symmetry applies; the fact that this is true for the physically motivated definition of $M^2=\diag(m_1^2,m_2^2,m_3^2)$ is one of convenience.

 {We also note that the form of the parameter symmetries as shown in section \ref{sec:summary of symmetries} and elsewhere through this paper is not fully symmetric under the different parameters.
This is because the specific description of parameter symmetries discussed depend on the parameterization of the mixing matrix.
A different parameterization in terms of a different sequence of rotations will result in similar parameterization symmetries where one needs to pay attention to which rotation is next to the diagonal mass matrix for the 1-2 interchange (which could become the 1-3 interchange or the 2-3 interchange accordingly) and which rotation is the final one to get to the flavor basis.}

 {At least some of the discrete  {parameter} symmetries can also be embedded in a group structure \cite{Zhou:2020iei}.}

While we do not know for sure if the list presented here contains all of the relevant  {parameter} symmetries of this form, we do believe that it is complete.

\section{Physical Consequences}
\label{sec:physcisConseq}
 {All physical oscillation observables must satisfy these parameter symmetries.
In this section we discuss a few important examples of how these parameter symmetries appear in physical observable variables and constrain the dependence on the parameters.
}

The CP violating term in appearance experiments proportional to the Jarlskog invariant \cite{Jarlskog:1985ht} is given by
\begin{align}
\Delta P_{\rm CP} \propto  & ~~ s_{23} c_{23}\, s_{13} c^2_{13}\, s_{12} c_{12}\sin (\delta_{23}+\delta_{13}+\delta_{12}) \notag \\[2mm]  &\times \sin \Delta_{21} \sin \Delta_{31} \sin \Delta_{32}\,,
\label{eq:PCPV}
\end{align}
which is unchanged under all of the above  {parameter} symmetries  {including the CPT symmetry}.
We note that the fact that $m_{13}$ (the change on $c_{13}$) is treated differently from the other $m_{ij},n_{ij}$ is the same reason that $c_{13}^2$ appears in eq.~\ref{eq:PCPV}; if $m_{13}=1$ then $c_{13}^2\to c_{13}^2$ and thus we must have $\sin\delta\to\sin\delta$ to remain invariant while for the remaining $s_{ij}$ and $c_{ij}$ terms as each picks up a minus sign it is exactly offset by the minus sign in $\sin(\delta_{23}+\delta_{13}+\delta_{12})$.
This is a useful non-trivial cross check of the symmetries discussed in this paper.

The Hamiltonians are invariant under these  {parameter} symmetries as described above, but certainly individual elements that appear in the Hamiltonian are not, such as $s_{13}$ or $\Delta m^2_{21}$.
That said, there are a number of interesting non-trivial terms that appear regularly in exact and various approximate oscillation probabilities that are invariants of all of the  {parameter} symmetries.
We list the  {parameter} symmetry invariant factors here:
\begin{itemize}
\item $s_{13}e^{i\delta_{13}}$, $s^2_{13}$ and $c^2_{13}$,
\item $s_{23}c_{23}e^{i\delta_{23}}$, $s^2_{23}$ and $c^2_{23}$,
\item $s_{12}c_{12}e^{i\delta_{12}}\Delta m^2_{21}$,
\item $\cos2\theta_{12}\Delta m^2_{21}=(c_{12}^2-s_{12}^2)\Delta m^2_{21}$,
\item $c_{12}^2\Delta m^2_{31}+s_{12}^2\Delta m^2_{32}\equiv\Dmsqee$ (see ref.~\cite{Nunokawa:2005nx}).
\end{itemize}
In addition, combinations of these expressions appear in the probabilities.
For example, the usual $\cos\delta$ or $\sin\delta$ terms  {must} appear in the combination, $s_{13} s_{23}c_{23} s_{12}c_{12}e^{i(\delta_{23}+\delta_{12}+\delta_{13})}\Delta m^2_{21}$ possibly with additional  {parameter} symmetry invariant factors such as $c_{13}^2$, $s^2_{23}$, \dots.
See, for example, table 1 of \cite{Denton:2016wmg}.
 {Also, these parameter symmetries exclude some combinations of parameters in physical observables, such as, odd powers of $c_{13}$.}
 
\section{ {Parameter} symmetries of approximations in the literature}
\subsection{Simple probability approximations}
The DMP approximation \cite{Denton:2016wmg} has many useful features including the fact that it automatically respects the maximum number of  {parameter} symmetries for an approximation scheme.
There are numerous other interesting approximate expression in the literature, see \cite{Parke:2019vbs} for a review.
While many of these approximate expressions do not follow the form of eq.~\ref{eq:Hmatpert}, we nonetheless investigate which of these  {parameter} symmetries they respect.

We first note that all other expressions only consider the single complex phase $\delta\equiv\delta_{23}+\delta_{13}+\delta_{12}$, thus the delta shuffle  {parameter} symmetry is not relevant.
Next, we investigate the behavior of these approximations under the discrete  {parameter} symmetries.

We begin with the expressions for the probabilities written as simple functions of the vacuum parameters and the matter potential.
As such, there are no approximate matter  {parameter} symmetries.
This includes 7 expressions\footnote{We consider eqs.~31 and 48 of \cite{Akhmedov:2004ny} and eq.~36 of \cite{Freund:2001pn}.} \cite{Cervera:2000kp,Akhmedov:2004ny,Friedland:2006pi,Arafune:1997hd,Freund:2001pn,Asano:2011nj}.
All of these expressions respect the $2^6=64$  {parameter} symmetries of the mixing angles represented by the $m_{ij},n_{ij}$ except for that of ref.~\cite{Freund:2001pn} which only respects the  {parameter} symmetry associated with $m_{13}$.
On the other hand, \emph{none} satisfy the 1-2 interchange, although many could with a simple change of $\Delta m^2_{31}\to\Dmsqee$ (a change which, in some cases, is known to increase the precision of approximations \cite{Parke:2016joa,Denton:2019yiw}).
 {In addition, since these are all expressions as functions of the vacuum parameters only but include the matter effect, none satisfy the CPT symmetry of the vacuum parameters either.}

 {One interesting approximate expression is that from \cite{Cervera:2000kp} which we reproduce here\footnote{The notation is slightly modified and the absolute value signs are removed as they are not relevant, for convenience.},
\begin{align}
P_{\mu e}={}&4s_{23}^2s_{13}^2c_{13}^2\left(\frac{\Delta m^2_{31}}{\Delta m^2_{31}-a}\right)^2\sin^2\left(\frac{(\Delta m^2_{31}-a)L}{4E}\right)\nonumber\\
&+4c_{23}^2s_{12}^2c_{12}^2\left(\frac{\Delta m^2_{21}}a\right)^2\sin^2\left(\frac{aL}{4E}\right)\label{eq:madird}\\
&+8J_r\left(\frac{\Delta m^2_{21}}a \right) \sin\left(\frac{aL}{4E}\right)  \notag \\ 
&\hspace{-1cm} \times \left(\frac{\Delta m^2_{31}}{\Delta m^2_{31}-a} \right)\sin\left(\frac{(\Delta m^2_{31}-a)L}{4E}\right)\cos\left(\delta+\frac{\Delta m^2_{31}L}{4E}\right)\,,\nonumber
\end{align}}
where $J_r=s_{23}c_{23}s_{13}c_{13}^2s_{12}c_{12}$ is the reduced Jarlskog invariant \cite{Jarlskog:1985ht,Denton:2016wmg}.
We now rewrite this expression in such a way so that it respects the 1-2 interchange symmetry, and thus all the vacuum  {parameter} symmetries,
\begin{align}
P_{\mu e}={}&4s_{23}^2s_{13}^2c_{13}^2\left(\frac\Dmsqee{\Dmsqee-a}\right)^2\sin^2\left(\frac{(\Dmsqee-a)L}{4E}\right)\nonumber\\
&+4c_{23}^2c_{13}^2s_{12}^2c_{12}^2\left(\frac{\Delta m^2_{21}}a\right)^2\sin^2\left(\frac{aL}{4E}\right)\label{eq:sym prob}\\
&+8J_r \left( \frac{\Delta m^2_{21}}a \right) \sin\left(\frac{aL}{4E}\right) \notag \\ 
& \hspace{-1cm} \times \left(\frac{\Dmsqee}{\Dmsqee-a} \right) \sin\left(\frac{(\Dmsqee-a)L}{4E}\right)\cos\left(\delta+\frac{\Dmsqee L}{4E}\right)\,.\nonumber
\end{align}
Note that we have  {changed $\Delta m^2_{31}\to\Dmsqee$ and have also} added in a factor of $c_{13}^2$ to the second term compared to the expression in \cite{Cervera:2000kp}.
The impact of this $c_{13}^2$ term is small: 2\% on an already small term, but there is a slight improvement in the precision of the expression.
More importantly, it allows one to easily write eq.~\ref{eq:sym prob} as the sum of two amplitudes squared $P_{\mu e}=|\mathcal A_{\mu e}|^2$ where,
\begin{multline}
\mathcal A_{\mu e}=2c_{13}\left\{
s_{23}s_{13} \left( \frac\Dmsqee{\Dmsqee-a} \right)\sin\left(\frac{(\Dmsqee-a)L}{4E}\right)\right.\\
\left.+c_{23}s_{12}c_{12}\left(\frac{\Delta m^2_{21}}a \right)\sin\left(\frac{aL}{4E}\right)
\exp\left[i\left(\delta+\frac{\Dmsqee L}{4E}\right)\right]
\right\}\,.
\end{multline}

\subsection{Probability approximations with rotations}
The remaining approximate expressions \cite{Agarwalla:2013tza,Minakata:2015gra,Denton:2016wmg} use perturbative techniques of diagonalizing a part of the Hamiltonian.
In ref.~\cite{Agarwalla:2013tza} we note that the approximate eigenvalues denoted $\lambda'_\pm$ and $\lambda''_\pm$ do not respect the 1-2 interchange symmetry of the vacuum parameters, but do respect the $m_{ij},n_{ij}$ discrete  {parameter} symmetries of the vacuum angles.
With regards to the approximate matter variables, they work in the vacuum mass basis so the equivalent form of eq.~\ref{eq:Hmatpert} becomes
\begin{align}
H_{\rm flav}={}& U_{23} U_{13}U_{12}V_{12} V_{23}\Lambda V_{23}^\dagger V_{12}^\dagger U^\dagger_{12} U^\dagger_{13} U^\dagger_{23} \nonumber\\
 + & \left(U_{23} U_{13} U_{12}M^2U_{12}^\dagger U_{13}^\dagger U_{23}^\dagger+A  \right.  \\
&   \left.  -U_{23} U_{13}U_{12}V_{12}V_{23}\Lambda V_{23}^\dagger V_{12}^\dagger U^\dagger_{12}U^\dagger_{13}U^\dagger_{23}\right)\,, \notag
\label{eq:Hflav}
\end{align}
for some $V_{ij}$ and $\Lambda$.
Note that in \cite{Agarwalla:2013tza} the $U_{12}V_{12}$ rotations are subsequently combined into a single rotation.
In addition, the sequence of rotations is different for anti-neutrinos than for neutrinos.
Since the order of the diagonalization of the $V_{ij}$ matrices is not the same as the lepton mixing matrix, the discrete  {parameter} symmetries of the perturbative parameters in general don't remain.

Next, refs.~\cite{Minakata:2015gra,Denton:2016wmg} both start in the mass basis, perform a (23) rotation first with $\wt\theta_{23}=\theta_{23}$.
They then perform a (13) rotation for some angle $\wt\theta_{13}$.
After this \cite{Denton:2016wmg} also performs a (12) rotation for some angle $\wt\theta_{12}$ while \cite{Minakata:2015gra} is done and can be thought of as setting $\wt\theta_{12}=0$ and is thus of the same form for the context of the  {parameter} symmetries discussed here.
The remaining point to confirm is that the eigenvalues all satisfy the symmetries of the vacuum parameters which is shown in appendix \ref{sec:dmp}.
Thus the expressions in refs.~\cite{Minakata:2015gra,Denton:2016wmg} respect the $2^7$ discrete  {parameter} symmetries of the vacuum parameters and the additional $2^6$ discrete  {parameter} symmetries of the perturbative parameters provided that one considers $\wt\theta_{23}$ and $\wt\delta$ as separate parameters from $\theta_{23}$ and $\delta$ that are allowed to transform differently even though they take the same values in the standard ranges.
 {In addition, the CPT parameter symmetry of the matter parameters is also respected.}
Finally, we note that as pointed out in \cite{Parke:2019vbs} the approximation scheme in \cite{Denton:2016wmg} which respects all the possible  {parameter} symmetries of a perturbative expression is numerically more precise than the similar approach in \cite{Agarwalla:2013tza} which respects fewer  {parameter} symmetries.

\begin{figure*}
\centering
\includegraphics[width=0.49\textwidth]{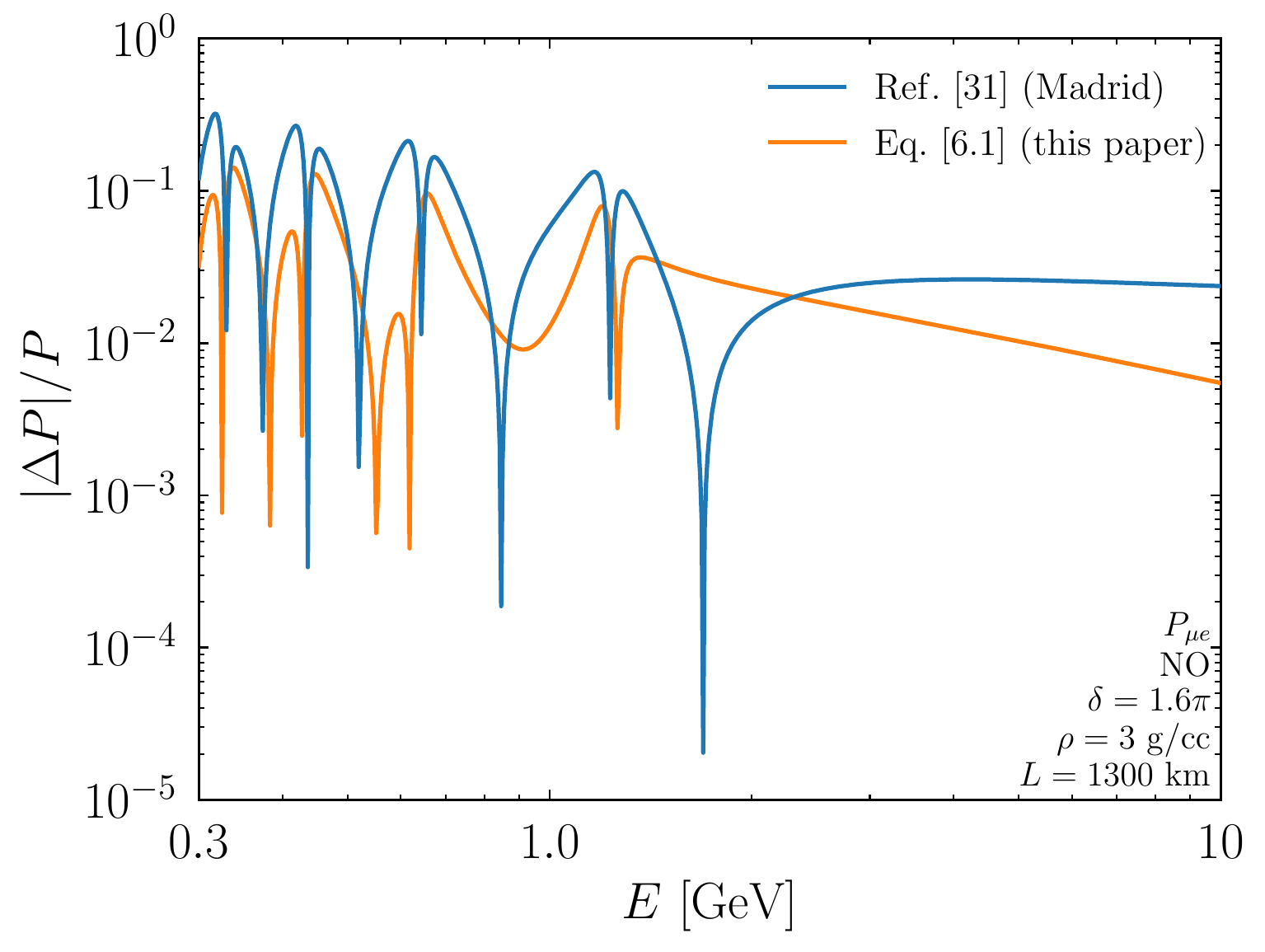}
\includegraphics[width=0.49\textwidth]{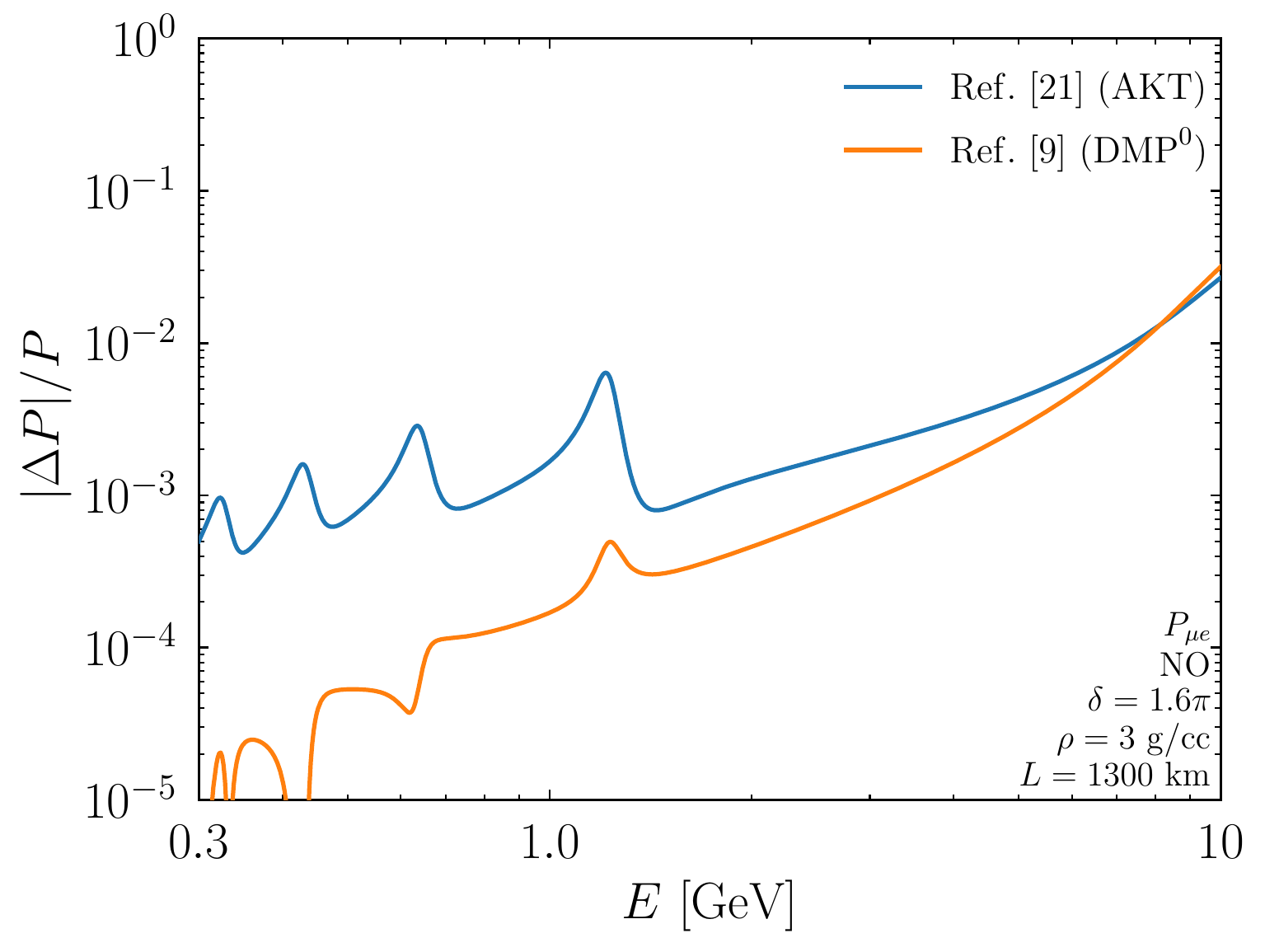}
\caption{Left panel: We show the fractional precision for the appearance probability in \cite{Cervera:2000kp} compared to the exact expression in blue.
In orange we change $\Delta m^2_{31}\to\Dmsqee$ which makes the expression in \cite{Cervera:2000kp} respect all $2^7=128$ discrete  {parameter} symmetries of the vacuum parameters including the 1-2 interchange and include a $c_{13}^2$ factor, see eq.~\ref{eq:sym prob}.
Right panel: we show the precision of AKT, \cite{Agarwalla:2013tza}, which does not respect the  {parameter} symmetries as well as DMP, \cite{Denton:2016wmg}, which does respect the  {parameter} symmetries. In both of these panels, expressions that respect the  {parameter} symmetries have better precision.
We use the best fit oscillation parameters from \cite{deSalas:2020pgw} and $\delta=1.6\pi$ to avoid any accidental CP phase cancellations.}
\label{fig:symmetry test}
\end{figure*}

In fact, we found that there is generally an increase in precision\footnote{The improved precision using $\Dmsqee$ instead of $\Delta m^2_{3i}$ broadly applies, except for a small range of $\delta$ values near $\delta=\pi/2$, a region that is disfavored by current T2K data \cite{Abe:2021gky}.} when using a form that respects all the available  {parameter} symmetries for both the simpler expressions discussed at the beginning of the section and those based on rotations in the previous papers.
In fig.~\ref{fig:symmetry test} we show the precision for the expressions from \cite{Cervera:2000kp,Agarwalla:2013tza,Denton:2016wmg} and eq.~\ref{eq:sym prob} above.
We also verified that in \cite{Akhmedov:2004ny,Friedland:2006pi} which are similar to \cite{Cervera:2000kp}, the original approximate expressions have a constant error $\sim2\%$ as $E\to\infty$, but a change from $\Delta m^2_{3i}\to\Dmsqee$ results in convergence at high energies for the expressions from \cite{Cervera:2000kp,Friedland:2006pi} and generally improved precision.

\subsection{ {Approximations for other oscillation parameters}}
 {
Another example of the impact of these parameter symmetries on the physics of neutrino oscillations is the size of the matter potential at the solar and atmospheric resonances given by
\begin{align}
a^R_{\text{sol}} & \approx \cos 2 \theta_{12} \Delta m^2_{21} / \cos^2 \theta_{13} \notag \\[2mm]
 \text{and} \quad a^R_{\text{atm}} &\approx \cos 2 \theta_{13} \Delta m^2_{ee}
\end{align}
respectively.
These expression are both extremely accurate as fractional corrections to these expressions are of order \cite{Parke:2020wha}
\begin{equation}
{\cal O}(s^4_{13}, ~s^2_{13} (\Delta m^2_{21}/\Delta m^2_{ee}),~ (\Delta m^2_{21}/\Delta m^2_{ee})^2)\,.
\end{equation}
The above expressions for the matter potential at the resonances satisfy all the parameter symmetries including the 1-2 symmetries.
Note for the atmospheric resonance it is $\Delta m^2_{ee}$, not $\Delta m^2_{31}$ or $\Delta m^2_{32}$ that appears, as the approximation is best when the 1-2 interchange symmetry is respected.}

 {The Jarlskog invariant in matter is the coefficient of the $\mathcal O(L^3)$ term in the appearance probability, and can also be simply approximated as \cite{Denton:2019yiw}
\begin{align}
\wh J&\approx\frac J{\mathcal S_\odot\mathcal S_{\rm atm}}\,,\\
\mathcal S_\odot&\equiv\sqrt{(\cos2\theta_{12}-c_{13}^2a/\Delta m^2_{21})^2+\sin^22\theta_{12}}\,,\\
\mathcal S_{\rm atm}&\equiv\sqrt{(\cos2\theta_{13}-a/\Dmsqee)^2+\sin^22\theta_{13}}\,.
\end{align}
Each of these terms respects all the symmetries, and if $\Dmsqee$ is changed to $\Delta m^2_{31}$ or $\Delta m^2_{32}$ which do not respect these parameter symmetries, the precision of the approximation is a factor $\gtrsim10$ times worse.}

 {
The effective $\Delta m^2$'s for disappearance experiments in vacuum are given by
\begin{align}
\Delta m^2_{ee} &= \cos ^2 \theta_{12} \Delta m^2_{31}+ \sin^2 \theta_{12} \Delta m^2_{31}\,,\\[2mm]
\Delta m^2_{\mu \mu } & \approx \sin ^2 \theta_{12} \Delta m^2_{31}+ \cos ^2 \theta_{12}\Delta m^2_{31} \notag \\
&+ 2 \sin \theta_{23} \cos \theta_{23} \sin \theta_{13} \sin \theta_{12} \cos \theta_{12} \notag \\ 
& \times ~\Delta m^2_{21} \cos(\delta_{12}+\delta_{13} +\delta_{23}) / \cos^2 \theta_{23}\,,\\[2mm]
\Delta m^2_{\tau \tau } &\approx \sin ^2 \theta_{12} \Delta m^2_{31}+ \cos ^2 \theta_{12} \Delta m^2_{31} \notag \\
&-2 \sin \theta_{23} \cos \theta_{23} \sin \theta_{13} \sin \theta_{12} \cos \theta_{12} \notag \\ 
& \times~ \Delta m^2_{21} \cos(\delta_{12}+\delta_{13} +\delta_{23}) / \sin^2 \theta_{23}\,,
\end{align}
for $\nu_e$, $\nu_\mu$ and $\nu_\tau$ disappearance respectively \cite{Nunokawa:2005nx}.
Each of these satisfies the symmetries given in section \ref{sec:symmetries} for (12), (13), and (23) sectors independently.}
 {Similarly, ref.~\cite{Zhou:2016luk} found that the eigenvalues in matter can be better approximated by using an atmospheric mass splitting of either $\Dmsqee$ or $\frac12(\Delta m^2_{31}+\Delta m^2_{32})$, both of which respect the above parameter symmetries, over other possible definitions which may not respect these parameter symmetries.}

 {
Finally, we note that the relevant exact and approximate two-flavor $\Delta m^2$ and mixing angle for $\nu_e$ disappearance in matter \cite{Denton:2018cpu},
\begin{align}
\Delta\wh{m^2}_{ee} & =\wh{m^2}_3-(\wh{m^2}_1+\wh{m^2}_2) \notag \\ &-[m^2_3-(m^2_1+m^2_2)]+\Dmsqee\,,\\
\Delta\wt{m^2}_{ee}& =\Dmsqee\mathcal S_{\rm atm}\,,   \text{ and }
\sin2\wt\theta_{13} =\frac{\sin2\theta_{13}}{\mathcal S_{\rm atm}}\,,
\end{align}
also respect all of the above symmetries, and that other possible forms that don't respect these symmetries are less precise.
}

\section{Conclusions}
The PDG \cite{Zyla:2020zbs} parameterization of the lepton mixing matrix of three rotations has been the de facto standard parameterization for neutrino oscillations for many years now.
This parameterization has many favorable phenomenological properties \cite{Denton:2020igp}, but it also leaves open many  {parameter} symmetries.
These  {parameter} symmetries relate two seemingly different sets of parameters to the same underlying physics, such as $\theta_{13}=8.5^\circ$ and $\theta_{13}=171.5^\circ$.
In this paper we elucidate what these  {parameter} symmetries are in the context of changing signs of $\sin\theta_{ij}$ or $\cos\theta_{ij}$ and adjusting the appropriate complex phase adjustment.
These  {parameter} symmetries then allow one to define the mixing angles as within a range spanning $\pi/2$ radians subject to certain restrictions, typically taken to be $\theta_{ij}\in[0,\pi/2)$.
In addition, these  {parameter} symmetries allow one to either fix $\Delta m^2_{21}>0$ or $\theta_{12}\in[0,\pi/4)$ depending on one's preference.

While the allowed ranges of the vacuum parameters have been previously identified, the framework presented here makes the connection to the  {parameter} symmetries manifest, which makes it clear that the same  {parameter} symmetries separately apply to the parameters in the presence of matter.
Thus in matter one has $2^{14}=16,384$  {parameter} symmetries including both vacuum and matter parameters.
In addition, if one uses an approximate perturbative scheme such as that in DMP \cite{Denton:2016wmg}, one has nearly all of the  {parameter} symmetries to any order in perturbation theory as well, $2^{13}=8,192$, subject to key matching conditions between the vacuum parameters and the approximate matter parameters.
 {Finally, we note that CPT invariance leads to another factor of two in the number of symmetries for each vacuum, matter, and approximate matter expressions.}
All combined this paper highlights  {49,408} discrete  {parameter} symmetries  {which apply to both neutrino and anti-neutrino oscillations} across the three different frameworks.
 {The CPT parameter symmetry combined with one of the other discrete parameter symmetries gives rise to the well known LMA-Light -- LMA-Dark degeneracy.}

These  {parameter} symmetries not only make it clear from where the restricted ranges on the parameters come from, they also provide a powerful tool when working with  {approximations for various physical quantities}.
While such approximate expressions need not satisfy any of the  {parameter} symmetries mentioned here to be useful, these symmetries do provide an important cross check and generally leads to improved precision.

We have focused on the standard PDG parameterization of the lepton mixing matrix, but the results presented here apply to different parameterizations containing a different sequence of rotations after straightforward modifications.
It may be interesting to explore connections to other parameterizations involving generators of SU(3) \cite{Merfeld:2014cha,Boriero:2017tkh,Davydova:2019aat}, four complex phases \cite{Aleksan:1994if}, five rotations and a complex phase \cite{Emmanuel-Costa:2015tca}, the exponential of a complex matrix \cite{Zhukovsky:2016mon}.

During the completion of this paper, ref.~\cite{Minakata:2021dqh} appeared on a related topic.
We note that the various vacuum, matter, and perturbative  {parameter} symmetries mentioned there are all covered in this paper.
We explicitly show the connection to our work for a representative example in appendix \ref{sec:HM}.

\acknowledgments
We acknowledge useful discussions with Hisakazu Minakata.
PBD is supported by the US Department of Energy under Grant Contract DE-SC0012704.
Fermilab is operated by the Fermi Research Alliance under contract no.~DE-AC02-07CH11359 with the U.S.~Department of Energy.
This project has received funding/support from the European Union's Horizon 2020 research and innovation programme under the Marie Sklodowska-Curie grant agreement No 860881-HIDDeN.

\newpage

\begin{widetext}

\appendix
\section{Transformations of the rotation matrices}
\label{sec:transformation}
In this appendix we give the effects of the transformations given in section \ref{sec:symmetries} on the rotation matrices $U_{23}$, $U_{13}$ and $U_{12}$, respectively.

The application of the  {parameter} symmetries to the (23)-rotation,
\begin{align}
\begin{pmatrix}
1 & &\\
&(-1)^m c & (-1)^n s e^{i(\delta+(m+n)\pi)} \\ 
&-(-1)^n s e^{-i(\delta+(m+n)\pi) } & (-1)^m c
\end{pmatrix}=
\begin{pmatrix}
(-1)^m\\
& 1 \\
&& 1
\end{pmatrix}
\begin{pmatrix}
1&&\\
& c & s e^{i\delta} \\
& -s e^{-i\delta} & c 
\end{pmatrix} (-1)^m \,.
\label{eq:sector23}
\end{align}
In this equation: $c=c_{23},s=s_{23}, \delta=\delta_{23}$ with $m=m_{23}, n=n_{23}$.

The application of the  {parameter} symmetries to the (13)-rotation,
\begin{align}
\begin{pmatrix}
(-1)^m c &\quad & (-1)^n s e^{-i(\delta+n\pi)} \\ 
& 1\\
-(-1)^n s e^{i(\delta+n\pi) } & & (-1)^m c 
\end{pmatrix}
= 
\begin{pmatrix}
(-1)^m\\
& 1 \\
&& 1
\end{pmatrix}
\begin{pmatrix}
c & & s e^{-i\delta} \\
&1&\\
 -s e^{i\delta} & & c
\end{pmatrix}
\begin{pmatrix}
 1 \\
& 1\\
&&(-1)^m\\
\end{pmatrix}\,.
\label{eq:sector13}
\end{align}
In this equation: $c=c_{13}, s=s_{13}, \delta=\delta_{13}$ with $m=m_{13}, n=n_{13}$.
The left (right) diagonal matrix commutes with $U_{23}$ ($U_{12}$ and also $M^2$).

The application of the  {parameter} symmetries to the (12) rotation without the 1-2 interchange,
\begin{align}
\begin{pmatrix}
(-1)^m c & (-1)^n s e^{i(\delta+(m+n)\pi)} & \\ 
-(-1)^n s e^{-i(\delta+(m+n)\pi) } & (-1)^m c &\\
&&1
\end{pmatrix}
&= 
\begin{pmatrix}
c & s e^{i\delta} \\ -s e^{-i\delta} & c \\
&&1
\end{pmatrix}
\begin{pmatrix}
(-1)^m\\
& (-1)^m \\
&& 1
\end{pmatrix}\,.
\label{eq:sector12}
\end{align}
In this equation: $c=c_{12}, s=s_{12}, \delta=\delta_{12}$ with $m=m_{12}, n=n_{12}$.

The application of the Symmetries to the (12) rotation with 1-2 interchange,
\begin{align}
&
\begin{pmatrix}
(-1)^m s & (-1)^n c e^{i(\delta+(m+n+1)\pi)} & \\ 
-(-1)^n c e^{-i(\delta+(m+n+1)\pi) } & (-1)^m s &\\
&&1
\end{pmatrix}
\begin{pmatrix}
0& 1\\
1&0 \\
&& 1
\end{pmatrix}
= 
\begin{pmatrix}
c & s e^{i\delta} \\ -s e^{-i\delta} & c \\
&&1
\end{pmatrix}
\begin{pmatrix}
-(-1)^m e^{i\delta} \\
& (-1)^m e^{-i\delta} \\
&& 1
\end{pmatrix}\,.
\label{eq:sector12x}
\end{align}
In this equation: $c=c_{12}, s=s_{12}, \delta=\delta_{12}$ with $m=m_{12}, n=n_{12}$.
The second matrix in eq.~\ref{eq:sector12x} performs the $m^2_1 \leftrightarrow m^2_2$ interchange.

Then, under all the  {parameter} symmetries given in section \ref{sec:symmetries}, we have  {the} symmetry transformation equation for Hamiltonian in the flavor basis:
\begin{multline}
U_{23} U_{13} U_{12}M^2U^\dagger_{12} U^\dagger_{13} U^\dagger_{23}
\Rightarrow \diag((-1)^{(m_{23}+m_{13})},1,1)U_{23} U_{13} U_{12}M^2U^\dagger_{12} U^\dagger_{13} U^\dagger_{23} \diag((-1)^{(m_{23}+m_{13})},1,1) \,.
\label{eq:mastereq}
\end{multline}
Only the diagonal matrices immediately after the equal side in eqs.~\ref{eq:sector23} and \ref{eq:sector13} appear in this transformation of the vacuum Hamiltonian in the flavor basis. This can also be seen directly by calculating the Hamiltonian in the flavor basis and then applying the  {parameter} symmetries. Only the first row and column get modified by multiplication by $(-1)^{(m_{23}+m_{13})}$. 
An identical result to eq.~\ref{eq:mastereq} exists for DMP at zero order, i.e.~for $V_{23} V_{13} V_{12}\Lambda V_{12}^\dagger V_{13}^\dagger V_{23}^\dagger $ with $(m_{23}+m_{13})$ replaced with $(\wt{m_{23}}+\wt{m_{13}})$. 

 {Adding the matter potential to both sides of eq.~\ref{eq:mastereq} does not effect this result as 
\begin{equation}
\diag((-1)^{(m_{23}+m_{13})},1,1)~ A ~\diag((-1)^{(m_{23}+m_{13})},1,1)=A\,.
\end{equation}
Therefore, eq.~\ref{eq:Htransformation} is valid in both vacuum and matter.}

\section{Delta shuffle examples}
\label{sec:shuffle explicit}
In this appendix we show exactly how one relates eq.~\ref{eq:Udeltas} to eqs.~\ref{eq:Udeltas23}-\ref{eq:Udeltas12} via diagonal rephasing matrices.
Recall that we drew attention to the ``delta shuffle'' which allows us to write the following mixing matrices in the same way
\begin{align}
U_{23}(\theta_{23},\delta_{23})U_{13}(\theta_{13},-\delta_{13})U_{12}(\theta_{12},\delta_{12})
&=D_{23}U_{23}(\theta_{23},\delta)U_{13}(\theta_{13},0)U_{12}(\theta_{12},0)D^\dagger_{23},\\
&=D_{13}U_{23} (\theta_{23},0) U_{13}(\theta_{13},-\delta)U_{12}(\theta_{12},0)D^\dagger_{13},\notag\\
&=D_{12}U_{23}(\theta_{23},0)U_{13}(\theta_{13},0)U_{12}(\theta_{12},\delta)D^\dagger_{12},\notag
\end{align}
where\footnote{Note that the $D_{ij}$ matrices may also have an additional overall phase which cancels out.}
\begin{align}
D_{23}=\diag(1,e^{-i\delta_{12}},e^{+i\delta_{13}}), \quad 
D_{13}=\diag(e^{+i\delta_{12}},1,e^{-i\delta_{23}}), \quad
D_{12}=\diag(e^{-i\delta_{13}},e^{+i\delta_{23}},1)
\end{align}
given that $\delta=\delta_{23}+\delta_{13}+\delta_{12}\mod\ 2\pi$.
Note, the $D_{jk}$'s are not unique.

\section{Parameter symmetries in MP}
\label{sec:MP}

In the approximation scheme of  \cite{Minakata:2015gra} (MP), where $V_{12}=1$, Eq. \ref{eq:Hmatpert} becomes
\begin{align}
H_{\rm flav} 
={}&\frac1{2E}\left[V_{23} V_{13} \Lambda  V_{13}^\dagger V_{23}^\dagger\right. 
   \left.+ (U_{23} U_{13} U_{12}M^2U_{12}^\dagger U_{13}^\dagger U_{23}^\dagger + A -V_{23} V_{13}  \Lambda  V_{13}^\dagger V_{23}^\dagger )\right]\,.    \label{eq:HmatpertMP}
\end{align}
Applying the $2^7$ vacuum symmetries to $\frac1{2E}\left[U_{23} U_{13} U_{12}M^2U_{12}^\dagger U_{13}^\dagger U_{23}^\dagger +A\right]$ results in the phase matrix $\diag((-1)^{m_{23}+m_{13}},1,1)$ multiplying 
the both from the left and the right, whereas $\frac1{2E}\left[V_{23} V_{13} \Lambda  V_{13}^\dagger V_{23}^\dagger\right]$ is invariant.

Applying the following symmetries for the matter variables
For $\wt{m}_{ij},\wt{n}_{ij}\in\{0,1\}$:
\begin{itemize}
\item $(23)$: $\wt{c}_{23} \rightarrow (-1)^{\wt{m}_{23}} ~\wt{c}_{23}$, $\wt{s}_{23} \rightarrow (-1)^{\wt{n}_{23}} ~\wt{s}_{23}$, and $\wt{\delta}_{23} \rightarrow \wt{\delta}_{23} \pm (\wt{m}_{23}+\wt{n}_{23}) \pi $,
\item $(13)$: $\wt{c}_{13} \rightarrow (-1)^{\wt{m}_{13}} ~\wt{c}_{13}$, $\wt{s}_{13} \rightarrow (-1)^{\wt{n}_{13}} ~\wt{ s}_{13}$, and $\wt{\delta}_{13} \rightarrow \wt{\delta}_{13} \pm (\wt{m}_{13}+\wt{n}_{13}) \pi$,
\item $(13)$: $\wt{c}_{13} \rightarrow (-1)^{\wt{m}_{13}} ~\wt{s}_{13}$, $\wt{s}_{13} \rightarrow (-1)^{\wt{n}_{13}} ~\wt{ c}_{13}$, and $\wt{\delta}_{13} \rightarrow \wt{\delta}_{13} \pm (\wt{m}_{13}+\wt{n}_{13} +1 ) \pi$,
plus $\lambda_- \leftrightarrow \lambda_+$,
\item The two phases given in $V_{23} (\wt{\theta}_{23},\wt{\delta}_{23}) V_{13} (\wt{\theta}_{13},-\wt{\delta}_{13}) $
are determined from the cancellations that must occur for the MP perturbation theory, these are:
\begin{align}
\wt{s}_{23}\wt{c}_{23}e^{i\wt\delta_{23}}&=s_{23}c_{23}e^{i\delta_{23}}\notag \\
\sign(\wt{c}_{23})\wt{c}_{13} \wt{s}_{13} e^{i\wt{\delta}_{13}} \Delta\lambda_{+-}&=\sign(c_{23})c_{13}s_{13}e^{i\delta_{13}}\Delta m^2_{ee} \,.
\end{align}
Not surprisingly, CP violation is determined by the sum of the vacuum $\delta$'s:  $\delta_{23}+\delta_{13}+\delta_{12}=\delta$ as in DMP. 
\end{itemize}
The above $2^5$ matter parameter symmetries give
\begin{align}
\frac1{2E}\left[V_{23} V_{13} \Lambda  V_{13}^\dagger V_{23}^\dagger\right]  \rightarrow  \diag((-1)^{\wh{m}_{23}},1,1) \frac1{2E}\left[V_{23} V_{13} \Lambda  V_{13}^\dagger V_{23}^\dagger\right]   \diag((-1)^{\wh{m}_{23}},1,1) \,.
\end{align}
 Matching the phase of $\nu_e$ between the two terms leads to the constraint that $\wh{m}_{23}=m_{23}+m_{13} $, so the total number of matter and vacuum parameter symmetries for MP perturbation theory is $2^{11}$.

\section{Symmetries of the exact Hamiltonian in matter}
\label{sec:zs}
The Hamiltonian in matter was first explicitly solved in \cite{Zaglauer:1988gz} (ZS), but the eigenvalues as written do not respect the  {parameter} symmetries, yet the probabilities must.
If we adjust the diagonal matrix by
\begin{equation}
M^2=(c^2_{12} m^2_1 +s^2_{12} m^2_2)\mathbb I+\diag(-s^2_{12} \Delta m^2_{21}, c^2_{12} \Delta m^2_{21}, \Delta m^2_{ee})\,,
\end{equation}
then coefficients of the characteristic equation,
\begin{equation}
(\wh{{m^2}_i})^3-A(\wh{{m^2}_i})^2+B\wh{{m^2}_i}-C=0\,,\label{eq:characteristic}
\end{equation}
where $\wh{{m^2}_i}$ are the three eigenvalues, are\footnote{ {$A$ and $C$ ($B$) are odd (even) under odd under the CPT symmetry, so that the $\wh{m_j^2}$'s are odd under this symmetry, as expected.} }
\begin{align}
A&=\Delta m^2_{ee} +(c^2_{12}-s^2_{12}) \Delta m^2_{21}+a\,, \\
B&=(c^2_{12}-s^2_{12})\Delta m^2_{21} \Delta m^2_{ee} -(s_{12} c_{12} \Delta m^2_{21})^2+a[c^2_{13}\Delta m^2_{ee}+(c^2_{12}-s^2_{12})\Delta m^2_{21}]\,,\\
C&=-(s_{12}c_{12}\Delta m^2_{21})^2\Delta m^2_{ee} +a[c^2_{13}(c^2_{12}-s^2_{12})\Delta m^2_{21}\Delta m^2_{ee}-s^2_{13}(s_{12}c_{12}\Delta m^2_{21})^2]\,.
\end{align}
From eq.~17 of ZS \cite{Zaglauer:1988gz}, the shift must be accounted for in order to see that the  {parameter} symmetries are respected.
Note that $\Delta m^2_{ee}$, $(c^2_{12}-s^2_{12})\Delta m^2_{21}$, and $(s_{12}c_{12}\Delta m^2_{21})^2$ are all unchanged under the 1-2 interchange.
Under this  {parameter} symmetry the mass-squared matrix changes
\begin{equation}
M^2 \Rightarrow \diag( -c^2_{12} \Delta m^2_{12}, s^2_{12} \Delta m^2_{12}, \Delta m^2_{ee}) = \diag( c^2_{12} \Delta m^2_{21}, -s^2_{12} \Delta m^2_{21}, \Delta m^2_{ee})\,,
\end{equation}
which leads to an identical characteristic equation, see eq.~\ref{eq:Msqs12sq}.

Recall that once the eigenvalues  {of the Hamiltonian and the eigenvalues of the principal minors of the Hamiltonian} are determined, the norm of the elements of the eigenvectors (or the norm of the elements of the diagonalizing matrix, $|W_{\alpha j}|^2$) are determined \cite{Denton:2019ovn,Denton:2019pka} and only the relative phases between the elements of the diagonalizing matrix are to be determined.  {There are multiple ways to choose these relative phases and this leads to most, but not all, of the parameter symmetries discussed in this paper.}

 {Alternatively, without using an explicit parameterization of the PMNS matrix, $U$, and using 
\begin{equation}
M^2=\diag (m^2_1,m^2_2, m^2_3)\,,
\end{equation}
then the coefficients of the characteristic equation (eq.~\ref{eq:characteristic}) are given by
\begin{align}
A={}& m^2_1+m^2_2+m^2_3+a  \\
B={}&m^2_1m^2_2+m^2_2m^2_3+m^2_3m^2_1 +a  \sum_i m^2_i (1-|U_{e i}|^2) \\
C={}&m^2_1m^2_2 m^2_3  
+a  \sum_{i,j,k} m^2_i m^2_j |U_{e k}|^2\,,     \quad \text{ ((i,j,k) all different)}
\end{align}
where $|U_{\alpha i}|^2$'s are the fraction of the $\alpha$-flavor in the i-th mass eigenstate in vacuum, and thus invariant under all the vacuum parameter symmetries of this paper. Therefore, the eigenvalues in matter, $\wh{m^2}_i$, are also invariant under the vacuum parameter symmetries.

In matter, the fraction of the $\alpha$-flavor in the i-th matter mass eigenstate is given by, see  \cite{Denton:2019ovn,Denton:2019pka},
\begin{align}
|W_{\alpha i}|^2 =& \frac{((\wh{m^2}_i)^2 -(\xi+\chi)_\alpha \, \wh{m^2}_i + (\xi \chi)_\alpha)}{(\wh{m^2}_i-\wh{m^2}_j)(\wh{m^2}_i-\wh{m^2}_k)}\,,
\end{align}
with (i,j,k) all different.
The sum and product of the eigenvalues of the principal minors are given by
\begin{align}
(\xi+\chi)_\alpha &= \sum_i m^2_i (1-|U_{\alpha i}|^2) 
+ \left\{ \begin{array}{ll}
 0 \quad & \alpha=e \\[2mm]
 a \quad & \alpha=\mu, ~\tau\
  \end{array}  
  \right.  
  \quad \text{with} \quad \sum_{\alpha} (\xi+\chi)_\alpha = 2A \,,
  \\[2mm]
(\xi\chi)_\alpha &= \sum_{i,j,k} m^2_i m^2_j |U_{\alpha k}|^2 
 + \left\{ \begin{array}{ll}
 0 \quad & \alpha=e \\[2mm]
 a \sum_i m^2_i|U_{\tau i}|^2  \quad & \alpha=\mu\\[2mm]
a  \sum_i m^2_i|U_{\mu i}|^2  \quad & \alpha=\tau
  \end{array}  
  \right.
   \quad \text{with} \quad \sum_{\alpha} (\xi\chi)_\alpha = B
 \,.
\end{align}
Thus, all vacuum parameter symmetries are respected by the matter  physical observables as well.  Only the unitarity of U has been used to derive these expressions\footnote{ {In particular, use of the identity $|U_{\alpha i}|^2= |U_{\beta j}U_{\gamma k}-U_{\beta k}U_{\gamma j}|^2$, with (i, j, k) and $(\alpha, \beta, \gamma)$ all different, is needed. This relation follows from $U^\dagger=U^{-1} = \text{adj}(U)/\text{det}(U)$.}}.
The variables $\wh{m^2}_i $'s and $|W_{e i}|^2 $'s are only dependent on $m^2_i$'s and $|U_{e i}|^2$'s and are explicitly  independent of  $|U_{\mu i}|^2$'s  and  $|U_{\tau i}|^2$'s ($\theta_{23}$ and $\delta$ in the PDG parameterization).

For oscillation physics, the only other quantity needed is the Jarlskog in matter, $\wh{J}$, which is related to Jarlskog in vacuum, $J$, by \cite{Naumov:1991ju,Harrison:1999df}
\begin{align}
\widehat{J} \, \Pi_{i>j} (\wh{m^2}_i-\wh{m^2}_j) &=  J  \, \Pi_{i>j}(m^2_i-m^2_j)\,.
\end{align}
As discussed earlier the LHS (RHS) of this equation are invariant under all matter (vacuum) parameter symmetries of this paper.
Thus, all physical observables of the exact solution in uniform matter, are  invariant under all parameter symmetries of both the matter variables and the vacuum variables. }

\section{Confirmation of the perturbative conditions for DMP}
\label{sec:dmp}
Ref.~\cite{Denton:2016wmg} (DMP) presented an approximation scheme for neutrino oscillations in matter based on diagonalizing ``most'' of the Hamiltonian via three rotations defined as a rotation of the vacuum parameters ($U_{23}(\theta_{23},\delta)$) and then diagonalizing a 2x2 submatrix twice in succession: 13 and then 12.
We here confirm that the requirements for the  {parameter} symmetries to hold are all satisfied by the details of this scheme.

For definiteness we consider NO $\then\lambda_1<\lambda_2<\lambda_3$.
We start with $M^2$ shifted as in eq.~\ref{eq:Msqs12sq}.
Then we define the initial eigenvalues after the (23) rotation before either of the diagonalizations,
\begin{equation}
\lambda_a= a+s^2_{13}\Dmsqee\,,\quad
\lambda_b=(c^2_{12}-s^2_{12})\Delta m^2_{21}\,,\quad
\lambda_c=c^2_{13}\Dmsqee\,,
\end{equation}
thus all  {parameter} symmetries satisfied for $(\lambda_a,\lambda_b,\lambda_c)$.

Next, after the (13) rotation we obtain the eigenvalues,
\begin{align}
\lambda_\pm &= \frac1{2} (a+ \Delta m^2_{ee}\pm \Delta \lambda_{+-})\,,\notag\\
\Delta \lambda_{+-} & =+\sqrt{(\Delta m^2_{ee} \cos 2 \theta_{13}-a)^2 + (\sin 2 \theta_{13} \Delta m^2_{ee})^2}\,,\\
\lambda_0 &= \lambda_b= (c^2_{12}-s^2_{12}) \Delta m^2_{21}\,,\notag
\end{align}
thus all  {parameter} symmetries are also satisfied for the $(\lambda_-, \lambda_0, \lambda_+)$ eigenvalues.
Therefore the diagonalizing angle, $\phi\equiv\wt\theta_{13}$, given by 
\begin{equation}
\sin^2 \phi=\frac{\lambda_+-\lambda_c}{\lambda_+-\lambda_-}\,,
\end{equation}
also satisfies all  {parameter} symmetries\footnote{The choice of sign when taking the square root gives rise to the  {parameter} symmetries.}.
This is the point at which \cite{Minakata:2015gra} stops, thus the eigenvalues in that paper all respect the  {parameter} symmetries of the vacuum parameters.

Continuing on with the framework of \cite{Denton:2016wmg}, after the (12) rotation,
\begin{align}
\lambda_3 &= \frac1{2} (a+ \Delta m^2_{ee}+ \sqrt{(\Delta m^2_{ee} \cos 2 \theta_{13}-a)^2 + (\sin 2 \theta_{13} \Delta m^2_{ee})^2})\,, \notag \\
\lambda_{2,1} &=\frac1{2} (\lambda_-+\lambda_0 \pm \Delta \lambda_{21} )\,,\\
\Delta \lambda_{21} &=+\sqrt{(\Delta m^2_{21} \cos 2 \theta_{12} -a_{12})^2 + (\cos^2\xi) ( \sin 2\theta_{12} \Delta m^2_{21})^2}\,, \notag
\end{align}
where  {parameter} symmetry invariant quantities $a_{12}$ and $\cos^2\xi$ are given by\footnote{In DMP, $\cos^2\xi$ is related to $\phi$ and $\theta_{13}$ via $\xi=\phi-\theta_{13}$ which may experience various reflections or shifts.}:
\begin{align}
a_{12} & =\frac1{2}(a+\Delta m^2_{ee} -\Delta \lambda_{+-})\,,\\
\cos^2\xi & = \frac1{2}\frac{\Delta m^2_{ee} +\Delta \lambda_{+-}-a\cos2\theta_{13}}{\Delta \lambda_{+-}}\,.
\end{align}
All  {parameter} symmetries satisfied for $(\lambda_1, \lambda_2,\lambda_3)$.

In the Hamiltonian whether we chose $\Lambda=\diag(\lambda_1,\lambda_2, \lambda_3)$ or $\Lambda=\diag(\lambda_2,\lambda_1, \lambda_3)$ determines whether the second diagonalizing angle, $\psi\equiv\wt\theta_{12}$, is given by
\begin{align}
\sin^2\psi\qquad{\rm or}\qquad\cos^2\psi=\frac{\lambda_2-\lambda_0}{\lambda_2-\lambda_1}\,.
\end{align}
Again we have the sign choices when the square root is taken.

Nothing here depends on what sign choices one has made in vacuum. However, whether $\delta_\phi =\delta_{13}$ or $\delta_\phi =\delta_{13} +\pi$ and similarly for $\delta_\psi$, does depend on both sets of choices.
This is related to eq.~\ref{eq:Hmatpert} where in DMP $V_{23}$ is a rotation of a new angle and phase that are taken to be the same as $\theta_{23}$ and $\delta$ in vacuum, $V_{13}$ is a rotation of the angle $\phi$, and $V_{12}$ is a rotation of the angle $\psi$.
This section confirms that the approximate eigenvalues that make up the $\Lambda$ matrix are invariant under all the  {parameter} symmetries as required by the second condition.

Here we precisely quantify how the procedure carried out in DMP \cite{Denton:2016wmg} is properly generalized to take advantage of all available  {parameter} symmetries.
First, we note that it makes sense to treat $\wt\theta_{23}$ as a separate parameter from $\theta_{23}$, just like how $\phi=\wt\theta_{13}$ and $\psi=\wt\theta_{12}$ are treated as separate variables from $\theta_{13}$ and $\theta_{12}$; similarly for the $\wt\delta_{ij}$.
Thus the conditions that need to be satisfied to follow the DMP procedure are
\begin{align}
s_{\wt{23}}c_{\wt{23}}e^{i\wt\delta_{23}}&=s_{23}c_{23}e^{i\delta_{23}}\notag\\
\sign(c_{\wt{23}})c_\phi s_\phi e^{i\delta_{\phi}}\Delta\lambda_{ee}&=\sign(c_{23})c_{13}s_{13}e^{i\delta_{13}}\Delta m^2_{ee}\label{eq:DMP diag conditions}\\
s_\psi c_\psi e^{i\delta_{\psi}}\Delta\lambda_{21}&=(|c_\phi c_{13}|+|s_\phi s_{13}|)s_{12}c_{12}e^{i\delta_{12}}\Delta m^2_{21}\,.\notag
\end{align}
From these equations its clear that the new phases must satisfy
\begin{align}
\wt\delta_{jk} & = \delta_{jk}\mod\ \pi\,.
\end{align}
In addition, the final requirement from section \ref{sec:perturbative},
\begin{align}
m_{13}+m_{23} & ={m_{\phi}}+\wt m_{23}\mod\ 2\,,
\end{align}
follows from the middle condition of eq.~\ref{eq:DMP diag conditions} when combined with the other requirements.

\section{Relationship to HM21}
\label{sec:HM}
Recently, \cite{Minakata:2021dqh} (HM21) appeared on a related topic.
All of the  {parameter} symmetries presented in HM21 fit within our framework.
To illustrate the relationship we pick one representative example containing all relevant features, Symmetry-IVB (last line of Table 1 in HM21) for which the  {parameter} symmetry is proven through first order in perturbation theory (our result in section \ref{sec:perturbative} is correct to all orders).

Symmetry-IVB is written as the following interchanges,
\begin{gather}
\theta_{23} \rightarrow - \theta_{23}\,,\qquad
\theta_{13} \rightarrow - \theta_{13}\,,\qquad
\theta_{12} \rightarrow- \theta_{12}\,,\qquad
\delta \rightarrow \delta + \pi\,,\nonumber \\
\phi \rightarrow - \phi\,,\qquad
\lambda_{1} \leftrightarrow \lambda_{2}\,,\qquad
c_{\psi} \leftrightarrow \pm s_{\psi}\,, ~~s_{\psi} \leftrightarrow \pm c_{\psi}\,.
\label{eq:Symmetry-V-DMP}
\end{gather}
We note that in the notation of HM21 (which matches that in DMP \cite{Denton:2016wmg} see also appendix \ref{sec:dmp}) $\lambda_i$ are the approximate matter eigenvalues of the perturbative matrix and $\phi$ ($\psi$) are $\wt\theta_{13}$ ($\wt\theta_{12}$).
Care is required in translating the statement of the  {parameter} symmetry in HM21 to that of this paper since in HM21 it is implicitly assumed that $\wt\theta_{23}$ is the same (and thus transforms the same) as $\theta_{23}$; $\wt\delta$ and $\delta$ are similarly linked.
In fact, each of the $\wt{\delta_{ij}}$ are taken to transform the same as the corresponding $\delta_{ij}$ (although this can be written as just a single complex phase for each side given the delta shuffle).

Then this  {parameter} symmetry with the upper signs is equivalent to our framework with the 1-2 interchange on the approximate variables, all the $m_{ij}=\wt m_{ij}=0$, and all the $n_{ij}=\wt n_{ij}=0$ except those listed here,
\begin{itemize}
\item for $\theta_{23} \rightarrow - \theta_{23}$, therefore $n_{23} =1$, this requires a $\pi$ be added to $\delta_{23}$,
\item for $\theta_{13} \rightarrow - \theta_{13}$, therefore $n_{13} =1$, this requires a $\pi$ be added to $\delta_{13}$,
\item for $\theta_{12} \rightarrow - \theta_{12}$, therefore $n_{12} =1$, this requires a $\pi$ be added to $\delta_{12}$,
\end{itemize}
thus $\delta \rightarrow \delta+\pi$ and $m_{23}+m_{13}=0$.
For the approximate matter parameters, we recall that we must also modify $\wt\theta_{23}$ and that the resultant modifications to $\wt\delta$ must match the factor of $\pi$ gained for $\delta$.
We have,
\begin{itemize}
\item for $\wt\theta_{23} \rightarrow - \wt\theta_{23}$, therefore $\wt n_{23}=1$, this requires $\pi$ be added to $\wt\delta_{23}$,
\item for $\wt\theta_{13} \rightarrow - \wt\theta_{13}$, therefore $\wt n_{13}=1$, this requires $\pi$ be added to $\wt\delta_{13}$,
\item $\wt{m^2_1} \leftrightarrow \wt{m^2_2}$ with $c_{\wt{12}} \rightarrow \pm s_{\wt{12}} $ and $s_{\wt{12}} \rightarrow \pm c_{\wt{12}} $ this requires $\pi$ be added to $\wt\delta_{12}$,
\end{itemize}
thus $\wt{\delta} \rightarrow \wt{\delta}+\pi$ and $\wt m_{23}+\wt m_{13}=0$.
So $\delta$ and $\wt\delta$ match as they must \emph{by the choice of definition for DMP} and this example satisfies our  {parameter} symmetry requirements including the constraint given by eq.~\ref{eq:m13m23pert}.
For the lower signs one additionally sets $\wt m_{12}=\wt n_{12}=1$, each of which results in a factor of $\pi$ added to $\wt\delta_{12}$ and thus no change to the complex phase.
We have also verified that our scheme contains all of the  {parameter} symmetries presented in table 1 of HM21.

Since HM21 uses the same symbol for $\delta$ in vacuum and in matter it important that they agree.
However, using the same symbol in vacuum and matter for $\theta_{23}$ and $\delta$ limits the number of possible discrete  {parameter} symmetries by $2^3=8$ (4 for $\theta_{23}$ and 2 for $\delta$).
Note that in a general perturbative framework, $\wt{\delta}$ and $\delta$ can differ by $\pi$.
Even accounting for these  {parameter} symmetries, there are still additional  {parameter} symmetries unexplored in HM21 such as those including a 1-2 interchange of both vacuum and matter parameters, those without the 1-2 interchange of the matter parameters, or those which send $c_{ij}\to-c_{ij}$ (and $c_{\wt{ij}}\to-c_{\wt{ij}}$).
Using a different symbol for $\theta_{23}$ in matter with its associated $\delta_{23}$ in matter, allows additional  {parameter} symmetries.
To address the maximum possible  {parameter} symmetries one must use different symbols for matter and vacuum for all variables and associate a separate $\delta$ for each angle, even though the oscillation probabilities only depend on the sum of these $\delta$'s.

\end{widetext}

\bibliography{Symmetries}

\end{document}